\documentclass[11pt,a4paper,twoside]{article}

\usepackage{axodraw}
\usepackage{psfrag}
\usepackage{epsfig}
\usepackage[latin1]{inputenc}
\def\be{\begin{equation}}
\def\ee{\end{equation}}
\def\ba{\begin{array}}
\def\bacc{\begin{array} {cc}}
\def\ea{\end{array}}
\def\bac{\begin{array} {c}}
\def\bea{\begin{eqnarray}}
\def\eea{\end{eqnarray}}
\def\bd{\begin{displaymath}}
\def\ed{\end{displaymath}}

\def\psl{\,\raise.15ex \hbox{/}\mkern-9.5mu \partial}

\begin{document}

\begin{center}

{\Large\bf Chiral Asymmetry from a 5D Higgs Mechanism}

\vspace{1cm}

Alberto Salvio\footnote{Email: alberto.salvio@epfl.ch} and Mikhail
Shaposhnikov\footnote{Email:
Mikhail.Shaposhnikov@epfl.ch}\\

\vspace{.6cm}

{\it { Institut de Th\'eorie des Ph\'enom\`enes Physiques,\\
  \'Ecole Polytechnique F\'ed\'erale de Lausanne,\\
  CH-1015 Lausanne, Switzerland}} \vspace{.4cm}
\vspace{.6cm}

\vspace{.4cm}

\vspace{.4cm}

\end{center}

\vspace{1.3cm}

\begin{abstract}

An intriguing feature of the Standard Model is that the
representations of the unbroken gauge symmetries are vector-like
whereas those of the spontaneously broken gauge symmetries are
chiral. Here we provide a toy model which shows that a natural
explanation of this property could emerge in higher dimensional
field theories and discuss the difficulties that arise in the attempt to construct a realistic theory.
An interesting aspect of this type of models is that
the 4D low energy effective theory is not generically gauge
invariant. However, the non-invariant contributions to the
observable quantities are very small, of the order of the square of
the ratio between the light particle mass scale and the Kaluza-Klein
mass scale. Remarkably, when we take the unbroken limit both the
chiral asymmetry and the non-invariant terms disappear.

\end{abstract}

\newpage

\tableofcontents

\section{Introduction}
One of the main topics in the Physics-Beyond-the-Standard-Model is
the attempt to understand fermion representations. From this point
of view, models with extra dimensions are useful as they can lead to
chiral 4D effective theories, both in standard Kaluza-Klein (KK)
scenarios \cite{Manton:1981es} and in brane worlds
\cite{Rubakov:1983bb}. In some cases, in addition to a chiral
spectrum, one can also obtain tachyonic scalar fields
\cite{Dvali:2001qr} that can be candidates for the Higgs field in 4
dimensions. These properties are intriguing as a natural question is
whether or not there exists a more fundamental relation between the
chiral asymmetry, which we observe in our world, and the Higgs
mechanism.

The answer to such a question would also help us to understand the
structure of the Standard Model (SM), which actually possesses the
following very special feature: vector-like representations always
correspond to unbroken gauge symmetries, that is the electromagnetic
and the color gauge symmetries, whereas chiral representations are
associated to the broken part of the SM gauge group. Is this just a
coincidence or is there a deep reason?

In the present paper we argue that there could be an explanation for
that in the framework of higher dimensional models. More precisely
we construct a simple 5D model in which a 4D chiral asymmetry and
the Higgs mechanism are related in a way that, when we continuously
turn on the Higgs order parameter, a vector-like fermion
representation becomes chiral.

This property emerges quite naturally in theories where left-handed
and right-handed fermions are (dynamically) localized on two
different branes. In this type of constructions, one important
theoretical issue is the gauge invariance of the 4D effective
theory. Indeed, if a chiral asymmetry appears in the 4D effective
theory, it may happen to have a gauge anomaly. However, we leave the
study of anomalies for a future work and concentrate here on a
semiclassical approximation\footnote{Here semiclassical
approximation means that loop contributions have not been included
in deriving the 4D effective theory.}.

In fact, even the purely bosonic sector of the low energy theory can appear to be not gauge invariant.
We address this problem by considering a simple bosonic completion,
which includes an Abelian gauge field and a charged scalar. Our 5D
bosonic action also includes some {\it weight functions} that may
have their origin in warped compactifications of more fundamental
theories \cite{Shaposhnikov:2001nz,Rubakov:1983bz,Randall:1999vf}.
By considering a simple set up for the weight functions, we prove
that, in the case in which the 5D gauge symmetry is broken, the
action for the light modes generally cannot be written as a gauge
invariant action (with at most a spontaneous breaking of the gauge
invariance). However, we also show that the values of the observable quantities are
extremely close to the predictions of a gauge invariant theory. When
the 5D gauge symmetry is restored, the 4D effective theory acquires
an exact gauge symmetry and, remarkably, this happens when the
fermion representation becomes vector-like and a gauge anomaly
cannot appear.

Here, we also comment on the nature of the Higgs mechanism that is
used to achieve the main purpose of the paper: we clarify that, in
our model, it is not possible to study the Higgs mechanism directly
in the 4D effective theory and we provide some link to a previous
work on this topic \cite{Randjbar-Daemi:2006gf}. This seems a very
important point that one should keep in mind in performing
dimensional reductions, where it is usually assumed that tachyons in
the 4D effective theory can consistently trigger the spontaneous
symmetry breaking. Indeed, our 5D model represents a counter example
for such a procedure.

Finally, we discuss non-Abelian generalizations of the simple
aforementioned $U(1)$ model. In particular we discuss the
$SU(2)\times U(1)$ case and describe our mechanism in a model that
resembles the electroweak theory in the low energy limit.  However,
it is unclear if a complete phenomenological description of the
masses and the interactions of the observed particles can be
achieved in this type of models. Indeed, in the minimal set up that we consider in the paper the correct interactions of the Z vector
boson cannot be reproduced and it seems quite difficult to achieve realistic fermion and vector boson masses.

The paper is organized as follows. In Section \ref{basic} we explain
in more detail our basic idea and we study a 5D spinor field coupled
to a background gauge field and a domain wall configuration. Then,
in Section \ref{Bcompletion} we provide a simple bosonic completion
in which the gauge field is promoted to a dynamical field associated
to a (spontaneously broken) gauge symmetry. Therefore, we are ready
to turn in Section \ref{4Deff} to the study of the gauge invariance
of the 4D effective theory. In Section \ref{Extensions} we study the
non-Abelian extensions. Finally, in Section \ref{conclusions} we
provide the conclusions and some outlook. We leave to the appendix a
technical discussion on the gauge fixing.
%\newpage

\section{Basic Idea}\label{basic}

In this section we explain how a relation between the
Higgs mechanism and the low energy chiral asymmetry can occur. We would like to construct a model in which exact low energy gauge
symmetry corresponds to a vector-like spectrum whereas the broken
phase is associated to a chiral one. Here we show that this idea can
be implemented by starting with a higher dimensional model.

As a matter of fact the 4D gauge fields coming from dimensional
reductions have generally a constant wave function along the
internal dimensions\footnote{An exception is given by 4D gauge
fields corresponding to the isometry group of the internal space
\cite{Salam:1981xd}, but here we only consider 4D vectors descending
from higher dimensional gauge fields and orthogonal to the gauge
field background. In this case it can be proved, by using the
formalism of \cite{Randjbar-Daemi:2002pq},
 that they do not mix with
the spin-1 fields coming from gravity and that they have constant
profiles if the gauge symmetry is unbroken.}. In the presence of a
higher dimensional Higgs mechanism, these gauge fields become
massive and in general acquire a non-trivial profile

\vspace{-0.57cm}
\begin{figure}[h]%\psfrag{a}{$v$}
\includegraphics[width=5.5cm]{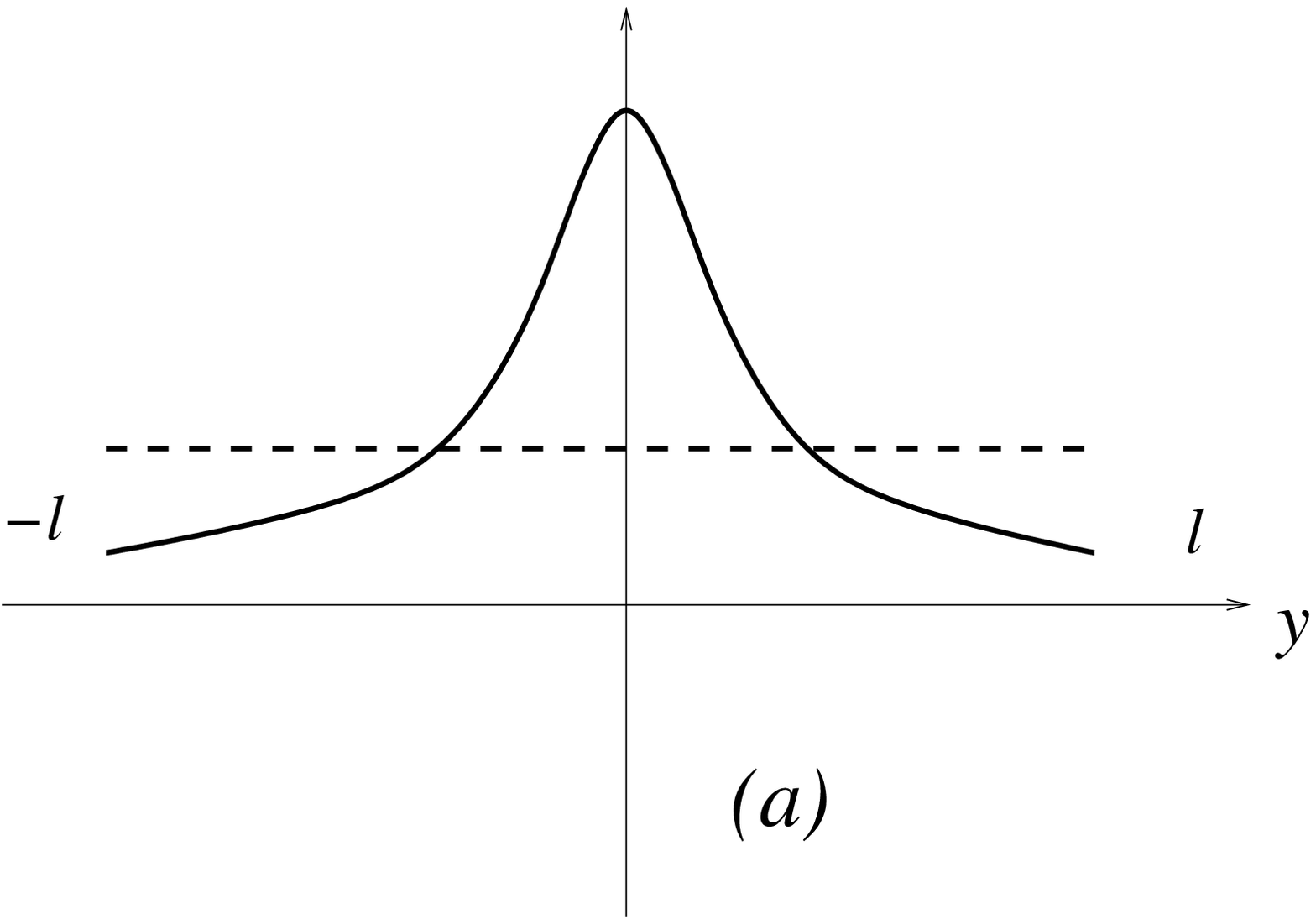}\hspace{2.3cm}%\vspace{0.3cm}
\parbox[b][5cm][t]{8cm}{
 \vspace{0.3cm} along the extra dimensions,
which can be peaked around some particular points. Now if fermions
with a given chirality, say left-handed fermions, are localized on
these points, whereas the right-handed modes are suppressed there,
the low energy theory will certainly present a chiral asymmetry.

\hspace{0.6cm} For the sake of definiteness, let us discuss now a
simple 5D case with an $S^1$ internal space. Therefore, the extra
dimension $y$ is subjected to the equivalence condition $y\sim
y+2l$, where $l/\pi$ is the radius of $S^1$. Without loss of
generality we focus on the region $-l\leq y \leq l$. In Plot $(a)$
we give the 4D vector field profile along the fifth dimension in two
cases: in the broken phase (continuous line) and in the unbroken
phase (dashed line). As we have pointed out, this function is
constant in the latter case, but is assumed to be peaked on some
point - here $y=0$ - in the former case. Then, in Plot $(b)$ we
present a simple step function domain wall configuration $\varphi$,
which can localize fermions with different chiralities on different
points of the fifth dimension \cite{Kaplan:1995pe, Fosco:1999bk};
this is actually a two domain wall configuration as the extra
dimension has a period $2l$. Finally, in Plot $(c)$, we show the
left-handed zero mode profile (continuous line) and the right-handed
one  \hspace{-0.4cm} }\vspace{0.2cm}
\\\includegraphics[width=5.5cm]{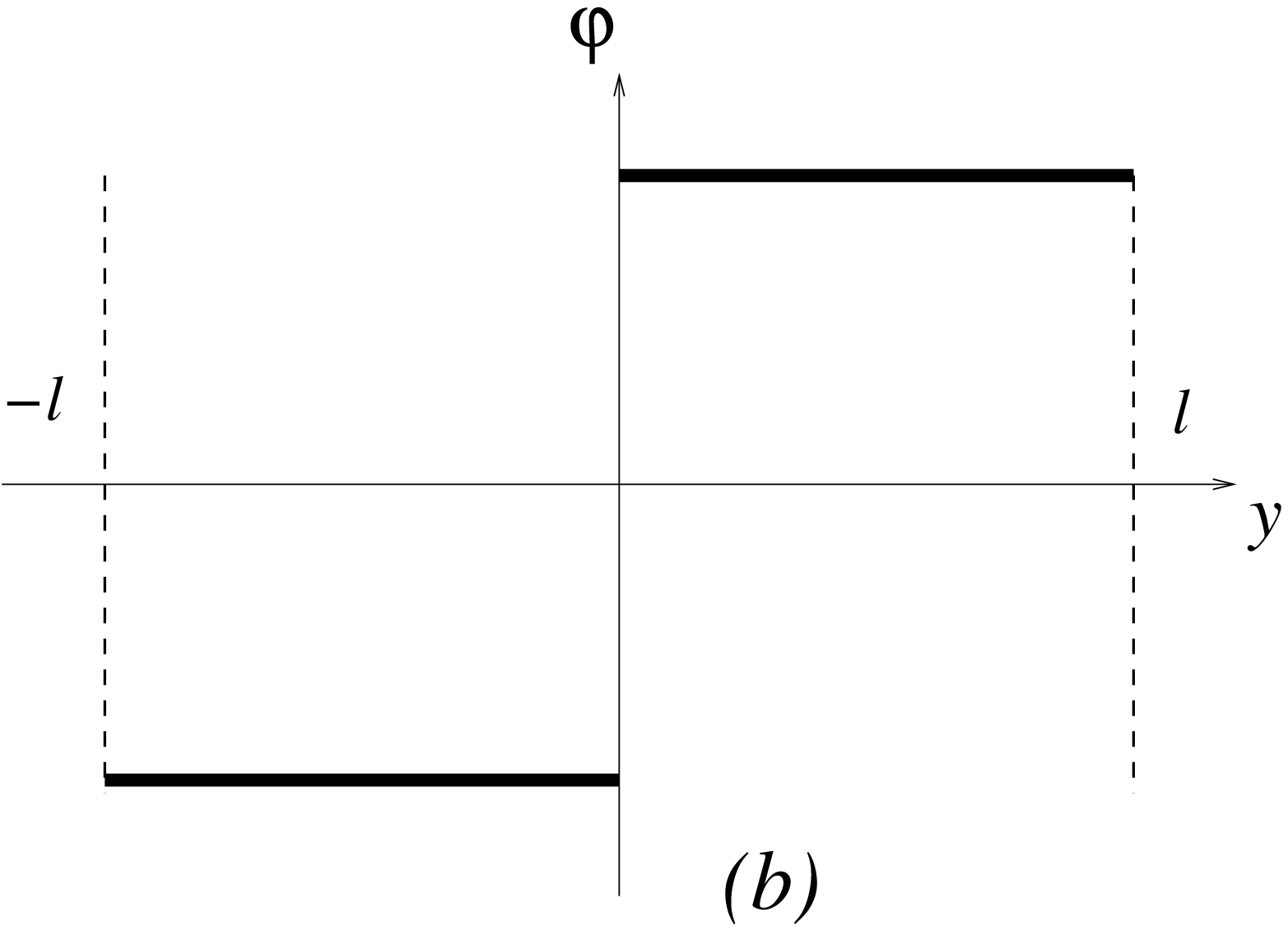}\vspace{0.2cm}
\\\includegraphics[width=5.5cm]{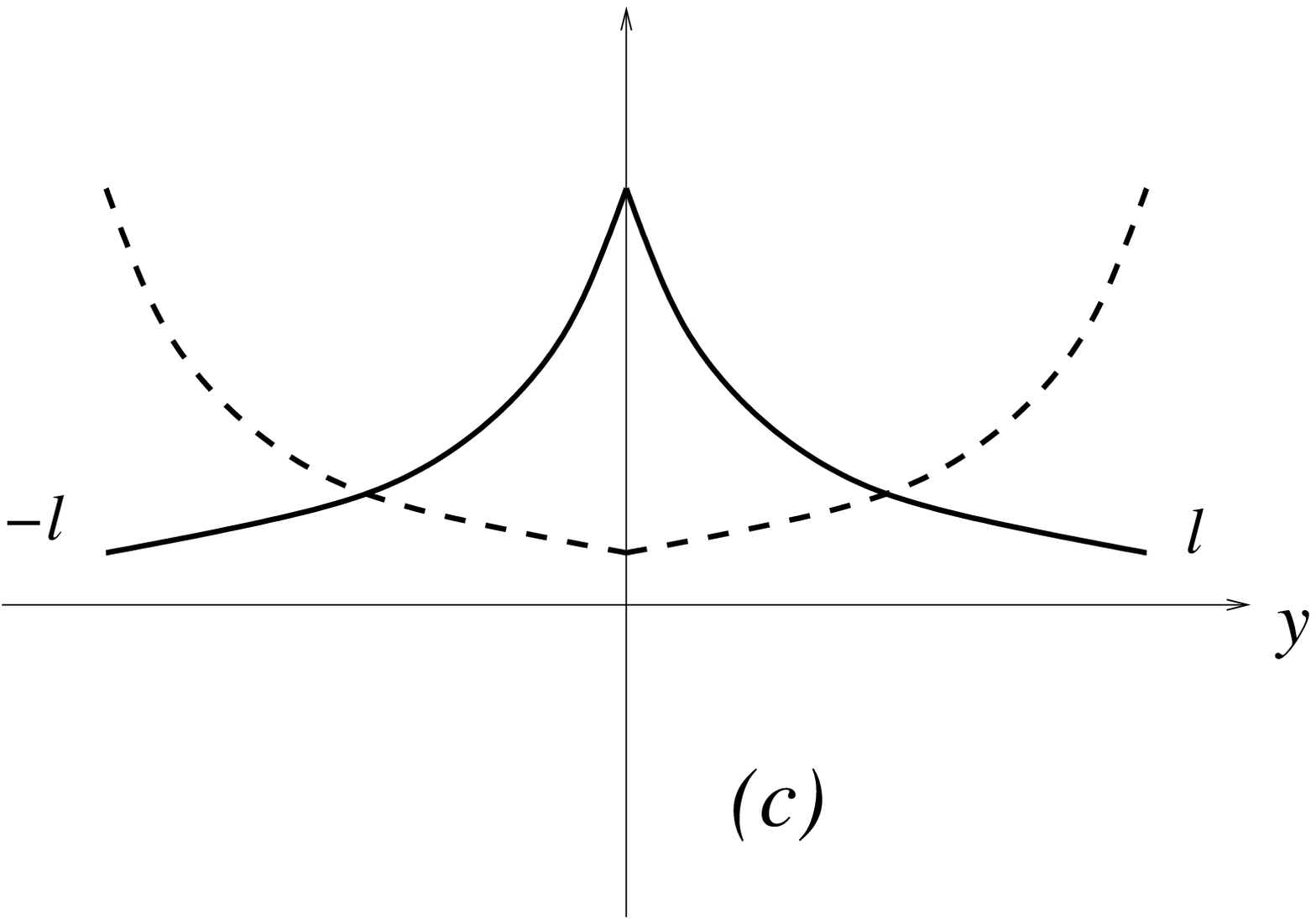}%\caption{Figure}%\label{fig1}
\end{figure}
\vspace{1cm}
\hspace{-0.7cm}     (dashed line), which Configuration (b) leads to. As we
can see, the left-handed mode is peaked on $y=0$, whereas the
right-handed mode is peaked on $y=\pm l$, which actually represents
a single point because the extra dimension is periodic. Since the
effective coupling constants between the fermions and the vector
field are given by the overlap integral between the function in Plot
$(a)$ and the absolute value squared of the fermion profiles in Plot
$(c)$, we expect a 4D chiral asymmetry in the broken case, which
should disappear when we take the unbroken limit.

 We will return to
this simple example in Subsection \ref{fermioncompact}, where we
will render the discussion more quantitative. Moreover, in
Subsection \ref{fermioninf} we shall prove that our basic idea can
be also implemented in the presence of an infinite extra dimension
and, therefore, is not based on the fact that the internal space is
compact.

Here we also stress that some ingredients that we used in this section
are well-known in the literature of higher dimensional model building: the fact that a non-trivial profile of the light gauge fields can emerge
from a Higgs mechanism and modify the 4D effective coupling constants have already been found \cite{Huber:2000fh}. As an original part here we establish that this
effect can actually be applied to find a relation between the Higgs mechanism and the low energy chiral asymmetry.

\subsection{Review of 5D fermions on domain walls}

In this subsection we review the basics of 5D fermions on domain walls in order to fix our conventions and to quantitatively implement our idea.

The simplest set up for the fermion action $S_F$ to realize our
mechanism consists of a 5D spinor $\psi$ and a domain wall
configuration $\varphi$, which traps 4D fermions with different
chiralities in different points of the fifth dimension. Therefore,
we write \cite{Rubakov:1983bb}
\be S_F=\int d^5 X\left(\overline{\psi}\,\Gamma^M D_M\psi +
\varphi\, \overline{\psi}\psi\right), \label{Faction}\ee
where $\varphi$ is assumed to be a non dynamical field, which
depends on the fifth dimension $y$, but is independent of the 4D
coordinate $x$, and $\Gamma^M$ are the 5D gamma matrices
($M=0,1,2,3,y$). Moreover, the covariant derivative $D_M\psi$ is
defined by $D_M\psi=(\partial_M+ie_fA_M)\psi,$ where $A_M$ is a
dynamical 5D gauge field and $e_f$ the corresponding fermion charge.
Action (\ref{Faction}) corresponds to the following equation of
motion (EOM):
\be \Gamma^MD_M\psi + \varphi\,\psi=0.\label{FEOM}\ee

We are now interested in the linear version of (\ref{FEOM}) with
respect to the dynamical fields, in order to extract information on
the fermion spectrum. By assuming the vacuum
expectation value (VEV) of $A_M$ equal to zero, this linearization
is
\be \left(\psl+\gamma^5\partial_y+\varphi \right)\psi=0,
\label{LFEOM}\ee
where $\psl\equiv\gamma^{\mu}\partial_{\mu}$ and we have introduced
the 4D gamma matrices $\gamma^{\mu}\equiv \Gamma^{\mu}$ and the 4D
chirality matrix $\gamma^5\equiv\Gamma^y$. We can now proceed in a
standard way and project (\ref{LFEOM}) onto the left-handed and
right-handed subspaces:
\be \psl\psi_R +\left(\partial_y + \varphi\right)\psi_L=0,\quad
\quad \psl\psi_L +\left(-\partial_y + \varphi\right)\psi_R=0,
\label{EOMLR}\ee
where we used $\gamma^5\psi_{L,R}=\pm \psi_{L,R}$. We want now to
study the 4D spectrum and, therefore, we perform a KK
decomposition as follows\footnote{In principle $n$ can be a discrete
or a continuous variable.}:
\be \psi_{L,R}(x,y)=\sum_n \psi^{(n)}_{L,R}(x)\, f_{L,R}^{(n)}(y).
\ee
From (\ref{EOMLR}) it is trivial to obtain and solve the equations
for the zero modes, which are defined by $\psl \psi_{L,R}^{(0)}=0$.
We have
\be f_L^{(0)}(y)\propto \exp\left[-\int^y dy' \varphi(y')\right],
\quad \,\,f_R^{(0)}(y)\propto \exp\left[+\int^y dy'
\varphi(y')\right]. \label{zeromodes}\ee
On the other hand, the equations for the massive modes, which satisfy
$\psl\psi^{(n)}_{L,R}=M_n\psi^{(n)}_{R,L}$, are
\be \left(-\partial_y^2+V_{L,R}\right)f^{(n)}_{L,R}=M^2_n
f^{(n)}_{L,R},\label{LRSchroedinger}\ee
which are Schroedinger equations with potentials $V_{L,R}=\mp
\partial_y\varphi + \varphi^2$. We can focus only on the
Schroedinger equation for one chirality, say the left-handed wave
functions $f^{(n)}_L$, because the right-handed counterpart can be
obtained by using the relation $M_n f^{(n)}_R = -(\partial_y +
\varphi)f^{(n)}_L$, which also follows from (\ref{EOMLR}). The
hamiltonians that appear on the left hand side of
(\ref{LRSchroedinger}) are hermitian. This property can be proved by
deriving the EOM from the action and by requiring that the boundary
terms, which come from the integration by parts, vanish.

\subsection{Simple example: 5D fermion on $S^1$}\label{fermioncompact}

We now return to the simple set up that we have qualitatively
discussed at the beginning of this section with the help of Plots
$(a)$, $(b)$ and $(c)$. As an original discussion, in this
subsection we quantitatively study how a low energy chiral asymmetry
and the spontaneous symmetry breaking (SSB) of the gauge symmetry
can emerge from a single mechanism in this simple case.

We observe that Plot $(b)$ corresponds to the configuration
$\varphi(y)=h(2\theta(y)-1)$, where $h$ is a positive constant and
$\theta(y)=1$ for $y>0$ and $\theta(y)=0$ for $y<0$. This domain
wall background leads to the following left-handed and right-handed
zero modes wave functions:
\bea
f_L^{(0)}(y)&=&\sqrt{\frac{h}{1-e^{-2hl}}}\,\,\exp\left\{\,h\left[\theta(-y)-\theta(y)\right]\,y\right\},\\
\nonumber
f_R^{(0)}(y)&=&\sqrt{\frac{h}{1-e^{-2hl}}}\,\,\exp\left\{\,-hl+h\left[\theta(y)-\theta(-y)\right]\,y\right\},
\eea
where we have used Eqs. (\ref{zeromodes}) and the normalization
constants have been computed in a way that the kinetic terms for
$\psi^{(0)}_{L,R}$ are canonically normalized. We notice that both
the zero modes satisfy the $S^1$ boundary conditions\footnote{We
remind that we have focused on the region $-l\leq y \leq l$ without
loss of generality.}, and that the discontinuity of $\varphi$ on
$y=0$ and  $y=l$ induces a discontinuity of the derivatives of both
the zero modes there.

It is now easy to see that the zero mode profiles have the shape
given in Plot (c). Moreover, the left-handed zero mode is
exponentially localized on $y=0$ and the right-handed one on $y=l$,
that is we have
\be
\frac{f_L^{(0)}(y=l)}{f_L^{(0)}(y=0)}=\frac{f_R^{(0)}(y=0)}{f_R^{(0)}(y=l)}=e^{-hl}.
\label{loczero}\ee
Eqs. (\ref{loczero}) tell us that the bigger is the dimensionless
parameter $hl$ the stronger is this localization mechanism.

As we pointed out at the beginning of this section, we now require
the lightest wave function $f_0(y)$ coming from $A_{\mu}(x,y)$ to
depend on some order parameter $v$ for the breaking of the 5D gauge
symmetry. Moreover, we require $f_0(y)$ to be peaked on $y=0$ for
$v\neq 0$ and to correctly reduce to a constant in the unbroken
limit ($v\rightarrow 0$). For example we could choose
\be f_0(y)=N_0 e^{-a^2(v)\, y^2}, \label{gaussian}\ee
with $a(v)\rightarrow 0$ in the unbroken limit. Wave function (\ref{gaussian}) is obviously of the form presented in Plot $(a)$. Now it is already
clear that we have a chiral asymmetry in the low energy theory for
$v\neq 0$. In order to discuss quantitatively this mechanism we give
the couplings between the light 4D fermions and the light 4D vector
field\footnote{Without loss of generality we have also assumed $\int
dy \left|f_0(y)\right|^2=1$.}:
\be g_L=e_f \frac{\int dy \left|f_L^{(0)}(y)\right|^2 f_0(y)}{\int
dy \left|f_L^{(0)}(y)\right|^2}, \quad g_R=e_f \frac{\int dy
\left|f_R^{(0)}(y)\right|^2 f_0(y)}{\int dy
\left|f_R^{(0)}(y)\right|^2},\ee
where the integrals are performed over the complete range of $y$ (in
this case $-l\leq y \leq l$). Therefore, for $v=0$, we have an
exact vector-like spectrum ($g_L= g_R$). But, in the case $v\neq 0$,
we have a big chiral asymmetry $g_L\gg g_R$ when the distance $l$
between the left-handed and the right-handed branes is large, for
fixed $h$ and $f_0$.

\subsection{Infinite fifth dimension}\label{fermioninf}

In the last subsection we have considered a compact extra dimension.
Here we show that an equivalent result can be obtained in the
presence of an infinite fifth dimension ($-\infty \leq y \leq
+\infty$). Indeed, since the fermion localization is achieved by the
field $\varphi$ and the remaining 4 dimensions are of course assumed
to be infinite, it is natural to have also an infinite fifth
dimension. Moreover, the set up that we present here
provides a non vanishing mass for the lightest fermions.

\begin{figure}
\centering
\epsfig{file=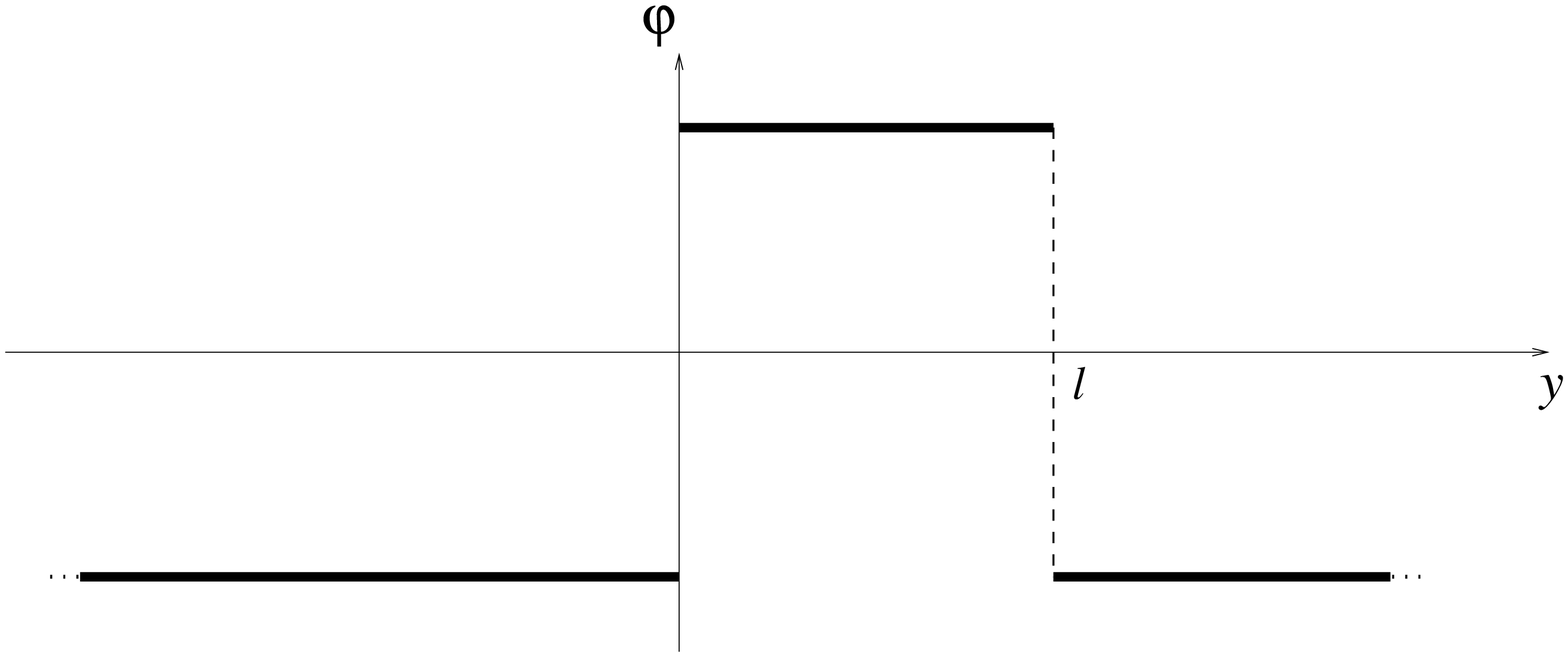,width=0.5\linewidth,clip=}\\\vspace{0.3cm}\hspace{0.1cm}\epsfig{file=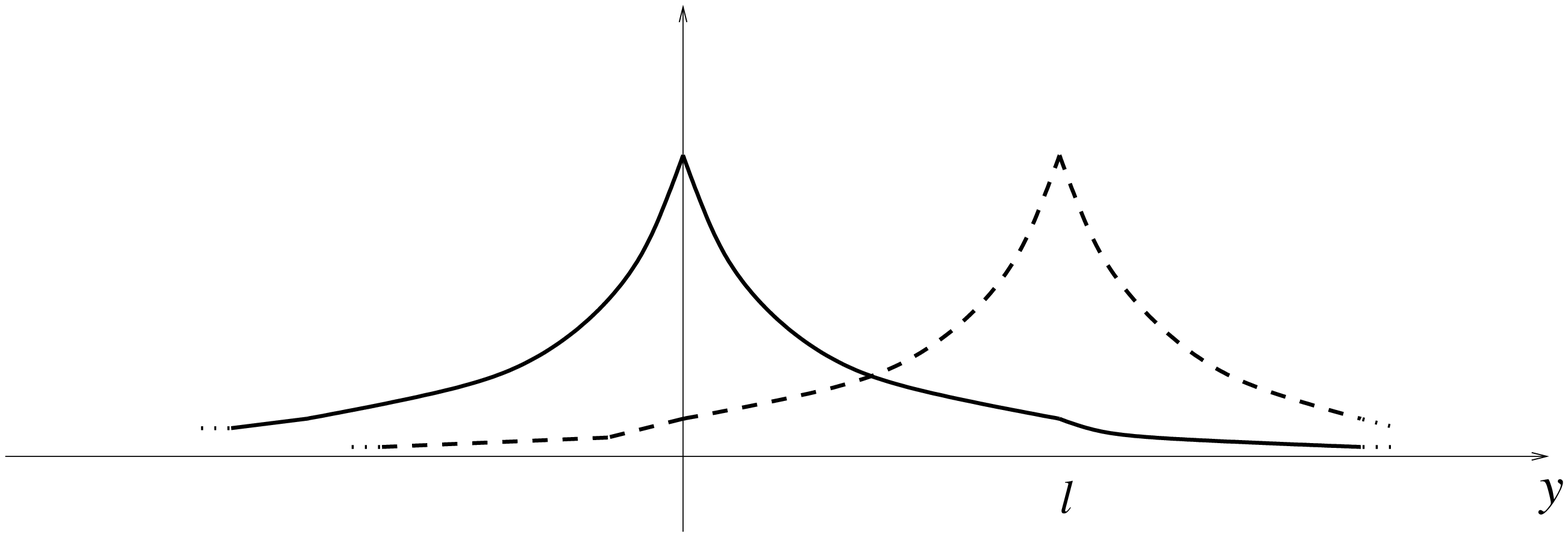,width=0.5\linewidth,clip=}
\caption{\footnotesize The first plot represents the two domain wall configuration defined by (\ref{phi2}). The second plot shows the corresponding
profiles of the left-handed (continuous line) and the right-handed (dashed line) lightest normalisable modes, given in Eq. (\ref{fLmassive}).}\label{infinite}
\end{figure}

In this subsection we consider the following configuration for
$\varphi$:
\be \varphi (y) = h\theta(y)\left[1-\theta(y-l)\right]-m, \label{phi2}\ee
where $h$ and $m$ are positive constants with dimension of mass and
$l$ is again a length scale. Eq. (\ref{phi2}), which corresponds to the first plot of Figure \ref{infinite}, represents the simplest example of a
two domain wall system.

By using (\ref{zeromodes}) we obtain the following solutions to the
zero mode equations:
\bea f^{(0)}_L(y) \propto &&  \exp\left[-\theta(y)hy+h\theta(y-l)\, (y-l) +my\right],\nonumber \\
f^{(0)}_R(y) \propto && \exp\left[+\theta(y)hy-h\theta(y-l)\,
(y-l) -my\right].\eea
These functions are both non normalizable and therefore they
decouple from the interactive sector of the effective theory. This
means that the lightest non-trivial mode is a couple of massive
(Weyl) fermions. Here we want to find explicitly their corresponding
mass and wave functions and so we consider the Schroedinger eqs.
(\ref{LRSchroedinger}) for the massive modes: in this case the
potentials for the two chiralities are
\bea V_{L,R} =&&
\left(h^2-2hm\right)\theta(y)\left[1-\theta(y-l)\right]+m^2\nonumber\\
           && \mp h \left[ \delta(y)-\delta(y-l)\right],\eea
where $\delta$ is the Dirac delta function which emerges from the
derivative of the step function $\theta$. We observe that the smooth
parts in $V_L$ and $V_R$ are equal. However, the left-handed modes
are subjected to a delta function with a negative coefficient at
$y=0$ and with a positive one at $y=l$, whereas, for the
right-handed modes, the delta functions are interchanged. Therefore,
we expect the lightest (massive) mode to be made of a left-handed
mode localized on $y=0$ and a right-handed mode localized on $y=l$.
To improve this localization mechanism we also impose $h^2-2hm>0$,
which, because of $h>0$, implies $h-m>0$.

By analyzing the Schroedinger equation for $f_L$ with standard
quantum mechanics methods, we find the following bounded solution:
\be \hspace{-0.1cm} f_L= \left\{\bac A_L \exp(qy) \quad \quad \mbox{for}\quad \quad
y<0,
\\\\
\frac{1}{2}A_L\left[ \left(1+\frac{q-h}{k}\right)\exp(ky)
+\left(1+\frac{h-q}{k}\right)\exp(-ky)\right] \quad  \mbox{for}
\quad 0\leq y\leq l,
\\\\
\hspace{-0.1cm}\frac{1}{2}A_L\left[ \left(1+\frac{q-h}{k}\right)\exp(kl)
+\left(1+\frac{h-q}{k}\right)\exp(-kl)\right] \exp(ql -
qy)\quad \mbox{for} \quad y>0,\ea \right.\label{fLmassive}\ee
where $q$ and $k$ are two positive\footnote{We observe that $q>0$ is
a necessary condition for normalisability.} constants defined by
\be q^2\equiv m^2-M^2, \quad k^2\equiv (h-m)^2-M^2, \ee
and $A_L$ is a normalisation constant. Moreover, $M$ represents the
mass of this solution, which should be computed by solving the
following algebraic equation:
\be \tanh(kl)=\frac{qk}{hm - q^2}. \label{massfermioneq}\ee
The corresponding solution for $f_R$ is given by $f_R(y)\propto
f_L(l-y)$ as the potentials satisfy $V_R(y)=V_L(l-y)$.

 We now want to compute explicitly $M$ and show that, for large values of
$l$, $f_L$ is localized on $y=0$ and $f_R$ is localized on
$y=l$. To this end we observe that $\tanh(kl)$ goes to $1$
with exponential velocity when $l\rightarrow \infty$ and, by
plugging this result into (\ref{massfermioneq}), we obtain
$M\rightarrow 0$. Therefore, we try a solution of the form
$M^2=M_0^2 \exp(-\alpha l)$, where $M_0$ and $\alpha$ are some
positive constants, namely a solution that goes exponentially to
zero when $l\rightarrow \infty$. By using this ansatz in Eq.
(\ref{massfermioneq}), we obtain
\be M^2 = 4 m^2
\left(1-\frac{m}{h}\right)^2\exp\left[-2(h-m)l\right] + ...\,,
\label{appmassfermioneq}\ee
where the dots are small corrections of the order
$\exp\left[-4(h-m)l\right]$. Therefore, if $l$ is large, we can take
just the first term in Eq. (\ref{appmassfermioneq}). Since we have
an explicit expression for $M$, we can now study the profile of
$f_{L,R}$ by plugging (\ref{appmassfermioneq}) into
(\ref{fLmassive}). The result is that $f_L$ is exponentially
localized on $y=0$ and $f_R$ is exponentially localized on $y=l$. In
fact we have
\be
\frac{f_L(y=l)}{f_L(y=0)}=\frac{f_R(y=0)}{f_R(y=l)}=\left(1-\frac{m}{h}\right)e^{-(h-m)l}+...\,,
\ee
where the dots now represent terms of the order
$\exp\left[-2(h-m)l\right]$. We provide the shape of $f_L$ and $f_R$
in the second plot of Figure \ref{infinite}.

We could proceed as in Subsection \ref{fermioncompact} and prove
that a big chiral asymmetry emerges in this case if $l$ is large
and if the gauge field $A_{\mu}$ satisfies the hypothesis that we
mentioned in Subsection \ref{fermioncompact}. In the next section we
shall discuss how these properties of the gauge field can be
generated in a dynamical way.

\section{A Bosonic Completion: 5D warped Higgs
Model}\label{Bcompletion}

In the last section we assumed $A_M$ to have the following
properties:
\begin{enumerate}
\item The spectrum of the 4D vector fluctuations is made of a light mode and a tower of heavy KK modes.
\item The light spin-1 mode wave function is peaked on the left-handed brane if and only if the gauge symmetry is spontaneously broken.\end{enumerate}
The main purpose of the present section is to obtain dynamically
these properties in a simple way.

To this end we introduce, apart from a 5D $U(1)$ gauge field $A_M$, a
charged Higgs field $\phi$ with a Mexican hat potential and we
assume the following bosonic action\footnote{We choose signature
$(-1,+1,+1,+1,+1)$.}:
\be S_B=\int d^5 X\left\{-\frac{\Delta}{4}\,F_{MN}F^{MN}- \Delta_S
\left[\left(D_M\phi\right)^{{\dagger}}D^M\phi +
V(\phi)\right]\right\},\label{SB}\ee
where $F_{MN}=\partial_MA_N-\partial_NA_M$, $D_M\phi\equiv \left(
\partial_M +ie A_M\right)\phi$, $V(\phi)\equiv
\lambda\left(|\phi|^2-v^2\right)^2$,  $e$ and $v$ are
real %non-vanishing
 constants, $\lambda$ is a positive constant and the {\it weight functions} $\Delta$
and $\Delta_S$ depend only on $y$. The $U(1)$ gauge symmetry acts on $A_M$ and $\phi$ as follows
\be A_M\rightarrow A_M-\partial_M\alpha,\qquad \phi\rightarrow e^{ie\alpha}\phi. \label{gaugetransf}\ee
We have introduced the $\Delta-$weights in order to have Property 1. Indeed, as we have
seen in Section \ref{basic}, the distance $l$ between the left-handed
and right-handed branes, and therefore the size of the fifth
dimension, should be large in order to achieve a big chiral
asymmetry. In this limit, the KK mass scale becomes very small in
standard KK scenarios, where one does not introduce any weight
function and assumes a compact internal space. Instead, by choosing
$\Delta$ and $\Delta_S$ in a suitable way, we can decouple the KK
mass gap and the volume of the internal space\footnote{Recently it
has been shown that, in the presence of warping, such a decoupling
may happen even if one takes into account the gravitational
backreaction \cite{massgaps}.}. In the following we shall take an
infinite extra dimension ($-\infty<y<+\infty$) as the localization
of the spin-1 and spin-0 fields can be obtained by means of the
weight functions\footnote{This set up is not
necessary and we could consider a compact internal space.}. Here we
do not describe the origin of the weight functions, but we observe
that they may arise for example from warped solutions of field
theories that include gravity \cite{Rubakov:1983bz,Randall:1999vf}.
For this reason we shall refer to the model defined by $S_B$ as a 5D
warped Higgs model.

The EOMs, which correspond to $S_B$, are
\bea &&\frac{1}{\Delta}\,\partial_M\left(\Delta \, F^{MN}\right)=ie\frac{\Delta_S}{\Delta}\left[\left(D^N\phi\right)^{\dagger}\phi-\phi^{\dagger}D^N\phi\right],\nonumber\\
   &&  \frac{1}{\Delta_S}D_M\left(\Delta_S D^M\phi\right)=\frac{\partial V}{\partial \phi^{\dagger}}. \eea
In the rest of the paper we assume the following VEV:
\be \left<A_M\right>=0,\quad \left<\phi\right>=v,\label{VEV}\ee
which is a solution to the EOMs and represents the simplest way to
realize the SSB of the gauge symmetry.

\subsection{Gauge fixing and perturbations}

We now want to study in details the linear perturbations around the
VEV (\ref{VEV}). To this end we first have to consistently fix a
gauge in order to focus only on the physical spectrum coming from
$A_M$ and $\phi=v+(\sigma+i\eta)/\sqrt2$. Therefore, we have to add
a gauge fixing term $\mathcal{L}_{GF}$ to the 5D lagrangian. We
choose\footnote{We observe that this gauge fixing leads to a
non-trivial ghost action. However, we do not analyze such a term
because, in the present paper, we only compute some observable
quantities in the semiclassical approximation, where there are no
ghost contributions.}
\be
\mathcal{L}_{GF}=-\frac{\Delta}{2}\left[\frac{1}{\Delta}\partial_M\left(\Delta
A^M\right)+\sqrt{2}ev\frac{\Delta_S}{\Delta}\eta\right]^2,
\label{GF} \ee
because, as we show below, in this gauge the spin-1 and spin-0
fluctuations do not mix at the bilinear level in the action. This
gauge is a generalization of the $R_{\xi}$ gauge (with $\xi=1$) to
warped 5D models\footnote{For a discussion on some generalizations
of the $R_{\xi}$ gauge in unwarped models see \cite{Muck:2001yv}.}.
In  the appendix %\ref{Rxi}
we demonstrate, with a perturbative argument, that (\ref{GF}) is a
legitimate gauge fixing term.

The complete bosonic action $S_B'=S_B+S_{GF}$, where $S_{GF}=\int
d^5 X \mathcal{L}_{GF}$, can be written in terms of the fluctuations
$A_M$, $\sigma$ and $\eta$ as follows:
\bea S_B'=\hspace{-0.5cm}&&\int d^5
X\left\{-\frac{\Delta}{2}\partial_MA_N\partial^MA^N
+\frac{\Delta}{2}\partial_y^2\ln\Delta \,A_y^2
-e^2v^2\Delta_S
A_MA^M-\frac{\Delta_S}{2}\partial_M\sigma\partial^M\sigma\right.\nonumber
\\
&& -\frac{\Delta_S}{2}\partial_M\eta\partial^M\eta-2\lambda
v^2\Delta_S\sigma^2 -e^2v^2\frac{\Delta_S^2}{\Delta}\eta^2+\sqrt2 ev
\Delta
\partial_y\left(\frac{\Delta_S}{\Delta}\right)A_y \eta
\nonumber \\&&+e\Delta_S A_M\left(\eta\partial^M\sigma -\sigma
\partial^M\eta\right)-\sqrt2 ve^2\Delta_SA_MA^M \sigma -\frac{e^2}{2}\Delta_SA_MA^M \left(\sigma^2+\eta^2\right)\nonumber \\
&&\left.-\sqrt2 v \lambda \Delta_S \sigma
\left(\sigma^2+\eta^2\right)-\frac{\lambda}{4}\Delta_S\left(\sigma^2+\eta^2\right)^2\right\},\label{gaugefixedS}\eea
where we have neglected boundary terms, that is all the terms of the
form $\int d^5 X \partial_M F$, with $F$ a functional of the fields,
and we have used that $\Delta$ and $\Delta_S$ depend only on $y$.
The first two lines in (\ref{gaugefixedS}) represent the bilinear
terms in the bosonic sector, whereas the third and fourth lines are
the interaction terms. It is now clear that the spin-1 field $A_\mu$
and the spin-0 fields $A_y$, $\sigma$ and $\eta$ do not mix at the
bilinear level. However, we have a non-trivial mixing between $A_y$
and $\eta$. We shall solve this problem in Subsection \ref{spin-0}.

From the 5D point of view, $\eta $ represents the would-be Goldstone
boson for the breaking of the U(1) gauge symmetry. This field
appears explicitly in the lagrangian because our gauge (\ref{GF}) is
not the unitary gauge, which instead corresponds to the $R_{\xi}$
gauge\footnote{See the appendix %\ref{Rxi}
for the definition of the
$R_{\xi}$ gauges in the presence of warping.} with $\xi\rightarrow
\infty$.

\subsection{Spin-1 sector and coupling with fermions}\label{spin-1}

In this subsection we focus on the spin-1 sector, which is crucial
in our discussion on the chiral asymmetry and the Higgs mechanism,
 and we study the linearized EOMs
for this sector.

To this end we need the bilinear action for $A_{\mu}$, which, thanks
to (\ref{gaugefixedS}), is simply
\be S_2(A_{\mu})=\int
d^5X\left(-\frac{\Delta}{2}\partial_{N}A_{\mu}\partial^{N}A^{\mu}-e^2v^2\Delta_SA_{\mu}A^{\mu}\right).
\ee
If we take the variation of $S_2(A_{\mu})$ under $A_{\mu}\rightarrow A_{\mu}+\delta A_{\mu}$ we obtain the following linearized EOMs:
\be \frac{1}{\Delta}\partial_M\left(\Delta
\partial^M A^{\mu}\right)=2e^2v^2\frac{\Delta_S}{\Delta}A^{\mu},\label{EOM1}\ee
where, as we did in the derivation of (\ref{gaugefixedS}), we have neglected
boundary terms. In the case of $A_{\mu}$, this constraint can be written as follows
\cite{Nicolai:1984jg,massgaps}:
\be \int dy\partial_y\left(\Delta \delta
A_{\mu}\partial_yA^{\mu}\right)=0.\label{HC1}\ee
Eq. (\ref{HC1}) represents a boundary condition for the wave
functions of the spin-1 fields along the extra dimension.

In order to analyze the 4D spectrum we perform a KK expansion,
$$A_{\mu}(x,y)=\sum_nA_{\mu}^{(n)}(x)f_n(y),$$
and a Fourier expansion of the 4D fields: $A_{\mu}^{(n)}(x)\propto
e^{-ip_nx}$. Eq. (\ref{EOM1}) now becomes
\be
-\frac{1}{\Delta}\partial_y\left(\Delta\partial_yf_n\right)+2e^2v^2\frac{\Delta_S}{\Delta}f_n=M^2_nf_n,\label{EOM2}\ee
where $M_n^2=-p_n^2$. The solutions to the latter equation, which
satisfy the boundary condition (\ref{HC1}), represent the physical
spin-1 sector that we are interested in. Eq. (\ref{EOM2}) can be
transformed in the standard Schroedinger form by means of the
definition $\chi_n\equiv \Delta^{1/2}f_n$:
\be
\left[-\partial_y^2+\mathcal{V}\right]\chi_n=M^2_n\chi_n,\label{V1}\ee
where the potential $\mathcal{V}$ turns out to be
$$\mathcal{V}=\frac{1}{4}\left(\partial_y\ln\Delta\right)^2+\frac{1}{2}\partial_y^2\ln\Delta+2e^2v^2\frac{\Delta_S}{\Delta}.$$
By expressing (\ref{HC1}) in terms of $\chi_n$, we also obtain
\be\int
dy\partial_y\left(\chi_{n'}\partial_y\chi_n-\frac{1}{2}\partial_y\ln\Delta\,
\chi_{n'}\chi_n\right)=0, \quad \forall\, n,\, n'.\label{HC2}\ee
 This boundary
condition implies the hermiticity of the hamiltonian in the
Schroedinger problem, and, therefore, we shall call it hermiticity
condition (HC) \cite{massgaps}.

So far we have considered general values of $\Delta$ and $\Delta_S$.
In the rest of this subsection we discuss the particular choice
\be \Delta(y)=\exp\left(-\frac{1}{2}\mathcal{M}^2y^2\right), \qquad
\Delta_S(y)=\frac{\delta^2}{8}y^2\exp\left(-\frac{1}{2}\mathcal{M}^2y^2\right),\label{choice}\ee
where $\mathcal{M}$ and $\delta$ are positive constants with the
dimension of mass. Indeed, as we prove below, (\ref{choice}) leads
to Properties 1 and 2 in a very simple way\footnote{Of course we do
not expect (\ref{choice}) to be the only set up that leads to those
properties, but we assume (\ref{choice}) for the sake of
definiteness.}. By plugging (\ref{choice}) into (\ref{V1}) we find
\be
\mathcal{V}(y)=\frac{1}{4}\mathcal{M}^4(1+\epsilon^2)y^2-\frac{1}{2}\mathcal{M}^2,\ee
where
\be \epsilon^2\equiv e^2v^2\delta^2/\mathcal{M}^4.\label{epsilon}\ee
Therefore, we have a harmonic oscillator potential. We observe that the only effect of
the SSB ($v\neq 0 $) is to change the ``frequency'' of the harmonic
oscillator as follows: $\mathcal{M}^2\rightarrow
\mathcal{M}^2\sqrt{1+\epsilon^2}\equiv \mathcal{M}^2_T$. It is now
trivial to obtain the spectrum, which is given by the harmonic
oscillator wave functions and ``energy'' eigenvalues:
\bea
\chi_n(y)&=&N_n\left(y-\frac{2}{\mathcal{M}^2_T}\,\partial_y\right)^n\exp
\left(-\frac{1}{4}\mathcal{M}_T^2y^2\right),\nonumber
\\ M^2_n&=&\mathcal{M}^2_T\left(n+\frac{1}{2}\right)-\frac{1}{2}\mathcal{M}^2, \quad n=0,1,2,...\,,\label{oscillator}\eea
where $N_n$ are normalization constants that can be fixed by
requiring standard kinetic terms for $A_{\mu}^{(n)}$. Moreover, it
is easy to see that the solutions given in (\ref{oscillator})
automatically satisfy (\ref{HC2}).

Therefore, the spectrum has the following features. We have a first
light mode with mass squared
$M^2_0=\frac{1}{2}\left(\mathcal{M}^2_T-\mathcal{M}^2\right)$, which
vanishes only for $v=0$. The corresponding wave function $f_0$ is
\be
f_0(y)=N_0\exp\left[-\frac{1}{4}\left(\mathcal{M}^2_T-\mathcal{M}^2\right)y^2\right].\ee
Hence, for $v=0$ we have a constant profile corresponding to a
massless gauge field, whereas in the broken case $v\neq 0$ the
lightest spin-1 field acquires a mass and it is localized on $y=0$
by means of a gaussian distribution of the form (\ref{gaussian}).
The remaining spin-1 states have very large masses if
$\mathcal{M}_T^2-\mathcal{M}^2 \ll \mathcal{M}^2$, which requires
$\epsilon^2\ll 1$. In this limit $\epsilon^2$ is of the same order
of magnitude as the ratio between the squared mass of the light
spin-1 field and the KK squared mass scale $\mathcal{M}^2$:
\be \epsilon^2\sim \frac{M^2_0}{\mathcal{M}^2}. \label{ordereps}\ee
So we find that Properties 1 and 2 are satisfied and, therefore,
(\ref{choice}) is a good choice to realize the mechanism that we
discussed in Section \ref{basic}.

Finally we note that, in order to obtain a large chiral asymmetry in
the broken phase, we need
$\left(\mathcal{M}_T^2-\mathcal{M}^2\right)l^2\gg 1,$ where $l$ is
the distance between the left-handed and the right-handed branes,
which we have introduced in Section \ref{basic}. Therefore, by using
$\epsilon^2\ll 1$, we find $\mathcal{M}^2\gg l^{-2}$, namely that a
decoupling between the KK mass scale and the size of the extra
dimension is needed, as we discussed at the beginning of this
section.

\subsection{Spin-0 sector}\label{spin-0}

We now complete the study of the linear perturbations around
(\ref{VEV}) by analyzing the spin-0 sector. Indeed, this is necessary
in order to know the complete low energy field content and if a
large mass gap between the light and heavy modes emerges in all
sectors.

We first examine the spectrum coming from $\sigma$, the 5D physical Higgs field. By using a method similar to what we have applied in the spin-1 sector,
we obtain the following linearized EOM and HC:
\be  \frac{1}{\Delta_S}\partial_M\left(\Delta_S\partial^M\sigma\right)=4\lambda v^2 \sigma,
\qquad\int dy \partial_y \left(\Delta_S \delta\sigma \partial_y \sigma \right)=0. \ee
After performing a KK decomposition
$\sigma(x,y)=\sum_n\sigma_n(x)f_{\sigma n}(y)$ and a Fourier
expansion of the 4D fields $\sigma_n(x)\propto e^{-ip_nx}$, again we
obtain a Schroedinger equation
\be
\left[-\partial_y^2+\mathcal{V}_{\sigma}\right]\chi_{\sigma n}=M^2_n\chi_{\sigma n},\quad \mathcal{V}_{\sigma}
=\frac{1}{4}\left(\partial_y\ln\Delta_S\right)^2+\frac{1}{2}\partial_y^2\ln\Delta_S+4\lambda v^2,\ee
where $\chi_{\sigma n}\equiv \Delta_S^{1/2}f_{\sigma n}$, and the following expression for the HC:
\be\int
dy\partial_y\left(\chi_{\sigma n'}\partial_y\chi_{\sigma n}-\frac{1}{2}\partial_y\ln\Delta_S\,
\chi_{\sigma n'}\chi_{\sigma n}\right)=0, \quad \forall\, n,\, n'.\label{HCsigma}\ee

If we consider Set up (\ref{choice}), the potential for $\sigma$ turns out to be
\be \mathcal{V}_{\sigma}=\frac{1}{4}\mathcal{M}^4y^2-\frac{3}{2}\mathcal{M}^2+4\lambda v^2, \ee
which is again a harmonic oscillator potential. Therefore, we find the following wave functions and masses squared:
\bea
\chi_{\sigma n}(y)&=&N_{\sigma n}\left(y-\frac{2}{\mathcal{M}^2}\,\partial_y\right)^n\exp
\left(-\frac{1}{4}\mathcal{M}^2y^2\right),\nonumber
\\ M^2_{\sigma n}&=&\mathcal{M}^2\left(n+\frac{1}{2}\right)-\frac{3}{2}\mathcal{M}^2+4\lambda v^2, \quad n=1,3,5,...\,.\label{oscillator2}\eea
% {ordereps}
The wave functions with $n$ even do not appear in (\ref{oscillator2}) because they do not satisfy the HC (\ref{HCsigma}).
 To illustrate this point let us take $n=0$ and $n'=0$ in the left hand side of (\ref{HCsigma}): we obtain
$$\int dy \partial_y\left(\chi_{\sigma 0}\partial_y\chi_{\sigma 0}+\frac{1}{2}\mathcal{M}^2y \chi_{\sigma 0}^2-\frac{1}{y}\chi_{\sigma 0}^2\right);$$
the first two terms in the previous expression vanish but the third
one does not because of the $1/y$ singularity, which is not canceled
by $\chi_{\sigma 0}^2$. Indeed, this result is not restricted to the
$n=0$ wave function, but it holds for all the wave functions with
$n$ even, as they are all  non-vanishing at $y=0$. Hence the set of
fluctuations emerging from $\sigma$ contains a light mode that
corresponds to $n=1$ in (\ref{oscillator2}) and a tower of KK modes.
The latter modes are much heavier than the $n=1$ mode when $\lambda
v^2 \ll\mathcal{M}^2$.

To complete the spin-0 sector we now have to examine the
fluctuations $A_y$ and $\eta$, which are coupled even at the
bilinear level. Indeed, the linearized EOMs for these fields are
\bea \frac{1}{\Delta}\partial_M\left(\Delta \partial^M A_y\right)&=&-\partial_y^2\ln\Delta\,A_y
+2e^2v^2\frac{\Delta_S}{\Delta}A_y-\sqrt2 e v \partial_y\left(\frac{\Delta_S}{\Delta}\right)\eta, \nonumber  \\
\frac{1}{\Delta_S}\partial_M\left(\Delta_S\partial^M\eta\right)&=&2e^2v^2 \frac{\Delta_S}{\Delta}\eta -\sqrt2 ev \partial_y\ln\frac{\Delta_S}{\Delta}A_y \label{EOMcoupled}
 \eea
and the corresponding HCs are
\be \int dy \partial_y\left(\Delta \delta A_y \partial_yA_y\right)=0,\qquad \int dy \partial_y\left(\Delta_S \delta\eta\partial_y \eta\right)=0.\label{HC3}\ee
We observe that Eqs. (\ref{EOMcoupled}) can be written in a
Schroedinger form by means of the transformations $\tilde{A}_y\equiv
\Delta^{1/2}A_y$ and $\tilde{\eta}\equiv \Delta_S^{1/2}\eta$:
\bea -\partial_y^2 \tilde{A}_y+\mathcal{V}_y \tilde{A}_y+\mathcal{C} \tilde{\eta}&=& \partial_{\mu}\partial^{\mu} \tilde{A}_y, \nonumber \\
 -\partial_y^2 \tilde{\eta}+\mathcal{V}_{\eta} \tilde{\eta}+\mathcal{C} \tilde{A}_y&=& \partial_{\mu}\partial^{\mu} \tilde{\eta}, \label{Schcoupled}\eea
where
\bea \mathcal{V}_{y}&=&\frac{1}{4}\left(\partial_y\ln\Delta\right)^2-\frac{1}{2}\partial_y^2\ln\Delta+2e^2v^2\frac{\Delta_S}{\Delta},\nonumber \\
\mathcal{V}_{\eta}&=& \frac{1}{4}\left(\partial_y\ln\Delta_S\right)^2+\frac{1}{2}\partial_y^2\ln\Delta_S+2e^2v^2\frac{\Delta_S}{\Delta},\nonumber \\
\mathcal{C}&=&-\sqrt2 ev \left(\frac{\Delta}{\Delta_S}\right)^{1/2}\partial_y\left(\frac{\Delta_S}{\Delta}\right).\eea
We observe that Eqs. (\ref{Schcoupled}) are also coupled and it is in
general difficult to find a complete set of solutions for general
$\Delta$ and $\Delta_S$. However, this problem can be easily solved
if we consider the special case given in (\ref{choice}). In fact, by
using (\ref{choice}) we obtain
\be \mathcal{V}_{\eta}=\frac{1}{4}\mathcal{M}_T^4
y^2-\frac{3}{2}\mathcal{M}^2,\quad
\mathcal{V}_y=\mathcal{V}_{\eta}+2\mathcal{M}^2,\quad
\mathcal{C}=-ev\delta. \ee
So $\mathcal{V}_{\eta}$ and $\mathcal{V}_y$ are two harmonic
oscillator potential and $\mathcal{C}$ is constant. These properties
allow us to easily decouple System (\ref{Schcoupled}). The mass
eigenstates can be expressed as follows:
\be \left(\bac \xi_1 \\ \xi_2\ea\right)\equiv \left(\bacc \cos \theta & \sin\theta \\ -\sin\theta & \cos\theta \ea\right) \left(\bac \tilde{A}_y \\ \tilde{\eta}\ea\right),
\label{xi12}\ee
where the mixing angle $\theta$ is defined by
\be \cos^2\theta\equiv \frac{\epsilon^2}{\epsilon^2+\left(\sqrt{1+\epsilon^2}-1\right)^2}. \ee
We recall that the parameter $\epsilon$, which we have defined in
(\ref{epsilon}), is very small because of (\ref{ordereps}). This
implies that $\theta$ is very small too: $\cos^2\theta =
1-\epsilon^2/4 +O(\epsilon^4)$. If we plug (\ref{xi12}) into
(\ref{Schcoupled}), we find two decoupled Schroedinger equations
with potentials $\mathcal{V}_{1,2}=\frac{1}{4}\mathcal{M}_T^4
y^2-\frac{1}{2}\mathcal{M}^2\pm \mathcal{M}_T^2$. The wave functions
and masses squared associated to $\xi_{i}$, $i=1,2,$ are
\bea
\chi_{i, n}(y)&=&N_{i, n}\left(y-\frac{2}{\mathcal{M}^2_T}\,\partial_y\right)^n\exp
\left(-\frac{1}{4}\mathcal{M}_T^2y^2\right),\nonumber
\\ M^2_{1(2),n}&=&\mathcal{M}_T^2\left(n+\frac{1}{2}\pm 1\right)-\frac{1}{2}\mathcal{M}^2, \quad n=1,3,5,...\,,\label{oscillator3}\eea
where again we have performed a KK decomposition
$\xi_i(x,y)=\sum_n\xi_{i,n}(x)\chi_{in}(y)$ and a 4D Fourier
expansion $\xi_{i,n}(x)\propto e^{-ip_{in}x}$. We observe that, as
in (\ref{oscillator2}), we do not have the even values of $n$ in the
harmonic oscillator spectrum because of the HCs. In fact, if we plug
(\ref{xi12}) and the definitions  $\tilde{A}_y\equiv
\Delta^{1/2}A_y$ and $\tilde{\eta}\equiv \Delta_S^{1/2}\eta$ into
(\ref{HC3}) we find
\bea \int dy \partial_y\left[\left(\cos\theta \, \delta\xi_1 -\sin\theta\, \delta\xi_2\right)\left(\partial_y-\frac{1}{2}\partial_y\ln\Delta\right)
\left(\cos\theta \, \xi_1-\sin\theta \, \xi_2\right)\right]&=&0,\label{HC4}\\
\int dy \partial_y\left[\left(\sin\theta \, \delta\xi_1 +\cos\theta\,\delta\xi_2\right)\left(\partial_y-\frac{1}{2}\partial_y\ln\Delta_S\right)
\left(\sin\theta \, \xi_1+\cos\theta \,\xi_2\right)\right]&=&0.\label{HC5}\eea
We now note that in (\ref{HC4}) we have $\partial_y\ln\Delta$, whereas in (\ref{HC5}) we have $\partial_y\ln\Delta_S$. Therefore, as
 we discussed below Eq. (\ref{oscillator2}),
Condition (\ref{HC4}) is weaker than (\ref{HC5}) and  we can just
focus on the latter. Also, if we first set $\xi_2=0$ and $\delta
\xi_2=0$ and we keep $\xi_1$ and $\delta\xi_1$ non vanishing and
then we exchange\footnote{This is possible because $\xi_1$,
$\delta\xi_1$, $\xi_2$ and $\delta \xi_2$ are all independent.} the
role of $\{\xi_1,\delta \xi_1 \}$ and $\{\xi_2, \delta \xi_2\}$, Condition (\ref{HC5}) implies
\be \int dy \partial_y\left[\delta \xi_i\left(\partial_y -\frac{1}{2}\partial_y\ln\Delta_S\right)\xi_i\right]=0. \ee
We can now apply the argument given below Eq. (\ref{oscillator2}) and find that only the odd values of $n$ are not projected out by the HCs.

So we have found that the sector $\{A_y,\eta\}$ contains just one
light mode ($\xi_{2,1}$, $i=2$, $n=1$) and a tower of heavy modes
with masses at least of the order $\mathcal{M}$.

\subsection{Counting the degrees of freedom}\label{counting}

\begin{table}[top]
\begin{center}
\begin{tabular}{|l|l|l|}
\hline  {\bf Sector} & {\bf ``Frequency''} & {\bf Squared masses}  \\ \hline
 $A_{\mu}$ & $\mathcal{M}^2_T$ &  $\mathcal{M}^2_T\left(n+\frac{1}{2}\right)-\frac{1}{2}\mathcal{M}^2$, $n=0,1,2,...$ \\
 $\sigma$ & $\mathcal{M}^2$ & $\mathcal{M}^2\left(n+\frac{1}{2}\right)-\frac{3}{2}\mathcal{M}^2+4\lambda v^2$, $n=1,3,5...$  \\
$\{A_y,\eta\}$ & $\mathcal{M}_T^2$ &
$\mathcal{M}^2_T\left(n+\frac{1}{2}\pm 1
\right)-\frac{1}{2}\mathcal{M}^2$, $n=1,3,5...$
 \\ \hline
\end{tabular}
\end{center}\caption{\footnotesize The complete bosonic spectrum in Case (\ref{choice}).
All the effective Schroedinger problems turn out to have a harmonic oscillator
potential with frequency given in the second column. The allowed levels in the harmonic oscillator spectrum are determined by the HCs.\label{spectrum}}
\end{table}

We conclude this section by summarizing the spectrum that we have found in the case of a  gaussian form (\ref{choice}) for the weight functions.
All the Schroedinger equations, which determine the mass spectrum and the profiles along the extra dimension, turn out to have a harmonic oscillator
potential. However, the ``frequencies'' of the oscillators and the boundary conditions are different in various sectors. We summarize
our results in Table \ref{spectrum}.

We observe that the low energy spectrum ($E\ll \mathcal{M}$)
contains a vector field with squared mass
$(\mathcal{M}_T^2-\mathcal{M}^2)/2$
 and only two scalar fields with squared masses $4\lambda v^2$ and $(\mathcal{M}_T^2-\mathcal{M}^2)/2$. Indeed, the field $\xi_1$
coming from the sector $\{A_y,\eta\}$ contains only heavy modes. We
conclude that the low energy spectrum and degrees of freedom are the
same as in the 4D spontaneously broken Higgs model in the $R_1$
gauge\footnote{In the $R_1$ gauge the mass of the vector boson is
equal to the mass of the would-be Goldstone boson.}.

In the next section we shall use the mass spectrum and the wave functions that we have found to study some interactions in the 4D effective theory.

\section{4D Effective Theory and Gauge Invariance}\label{4Deff}

In this section we study the form of the 4D effective theory for the
light bosonic modes. As we will see, when the 5D gauge symmetry is broken
($v\neq 0$), the action for such a theory in general cannot be
written  as a gauge invariant action with at most a SSB of the gauge
symmetry. We prove this statement by choosing the weight functions
(\ref{choice}) and exploiting the exact results of the previous
section.
 However, by using a semiclassical approximation, we also show that
the terms which explicitly break the 4D gauge invariance are very small (of the order of $\epsilon^2$) and,
therefore, in the limit in which the 5D gauge
symmetry  is restored ($v\rightarrow 0$) these terms go to zero and
the 4D effective theory acquires an exact gauge symmetry.

As stated in Subsection \ref{counting}, the low energy spectrum is
made of a vector boson $V_{\mu}\equiv A^{(0)}_{\mu}$ and two scalar
fields $\omega_1\equiv \sigma_1$ and $\omega_2 \equiv \xi_{2,1}$,
where the mass of $\omega_2$ is equal to the mass of $V_{\mu}$.
Therefore, if there was a gauge symmetry at low energy, the 4D
effective action would have the following form
\bea \hat{S}+\hat{S}_{GF}=\int d^4x&&\hspace{-0.5cm}\left[-\frac{1}{4}V_{\mu \nu}V^{\mu \nu}- \left(D_{\mu}\omega\right)^{{\dagger}}D^{\mu}\omega -U(|\omega|)\right.\nonumber \\
&&\left.-\frac{1}{2}\left(\partial_{\mu}V^{\mu}+\sqrt2 \hat{e}\hat{v}\omega_2\right)^2+...\right] ,\label{test}\eea
namely it should be the action of a 4D Higgs model in the $R_1$
gauge, apart from higher dimensional operators that we denoted in
(\ref{test}) with the dots\footnote{We note that the 4D effective
theory generally contains non renormalisable interactions and
therefore the theorem proved in \cite{Cornwall:1974km} is not
applicable.}.
 In (\ref{test}) we have introduced $V_{\mu \nu}\equiv \partial_{\mu}V_{\nu}-\partial_{\nu}V_{\mu}$, the complex field
$\omega\equiv \hat{v}+(\omega_1+i\omega_2)/\sqrt2$, the covariant derivative $D_{\mu}\omega=\left(\partial_{\mu}+i\hat{e}V_{\mu}\right)\omega$ and a general U(1)-invariant
potential $U(|\omega|)$. However, in this section we show that in general this is not the case when $v\neq 0$. To this end we explicitly write (\ref{test}) in terms of
$V_{\mu}$, $\omega_1$ and $\omega_2$:
\bea \hat{S}+\hat{S}_{GF}=\int d^4x\left[ -\frac{1}{2}\partial_{\mu}V_{\nu}\partial^{\mu}V^{\nu}-\hat{e}^2\hat{v}^2V_{\mu}V^{\mu}
-\frac{1}{2}\partial_{\mu}\omega_i \partial^{\mu}\omega_i-\hat{e}^2\hat{v}^2\omega_2^2\right.\nonumber \\
\left.+\hat{e}V_{\mu}\left(\omega_2\partial^{\mu}\omega_1-\omega_1\partial^{\mu}\omega_2\right)-\sqrt2 \hat{e}^2\hat{v}\omega_1V_{\mu}V^{\mu}
-\frac{1}{2}\hat{e}^2V_{\mu}V^{\mu}\left(\omega_1^2 +\omega_2^2\right)+U(|\omega|)+...\right],\label{test2}\eea
where we have taken $\hat{v}$ real without loss of generality. Since
$U$ depends only on $|\omega|$ it possibly contributes to the mass
term of $\omega_1$,
 but it gives no contribution to the mass term of $\omega_2$.

\subsection{Small explicit breaking of gauge invariance}

We now compare the 4D effective theory descending from the 5D theory of the previous section with the 4D theory defined by (\ref{test2}).
For simplicity we perform only semiclassical calculation,
that is we neglect all loop contributions to the effective theory.
Let us first ignore the higher dimensional operators that are represented by the dots in (\ref{test}); afterwards we will show that such operators
do not ruin the argument we are going to present here.

 If we compare the spectrum found in the previous section with the bilinear terms in (\ref{test2})
we obtain that gauge invariance requires
\be 2\hat{e}^2\hat{v}^2=\frac{1}{2}\left(\mathcal{M}_T^2-\mathcal{M}^2\right).\ee
This relation fixes one parameter out of $\hat{e}^2$ and
$\hat{v}^2$. The other parameter, say $\hat{e}^2$, can be fixed by
looking at the cubic operator
\be \omega_1V_{\mu}V^{\mu}. \ee
 In a realistic model an interaction of this type contributes to the decay of a physical Higgs into two massive vector bosons.
After a bit of algebra we obtain
\be \hat{e}^2=\frac{1}{2 \sqrt{2\pi}}\frac{e^2\mathcal{M}\epsilon^2}{(1+\epsilon^2)(\sqrt{1+\epsilon^2}-1)}.\ee
We can now test the gauge invariance of the 4D effective theory by examining the operator
\be V_{\mu}V^{\mu}\omega_1^2.\label{4operator}\ee
 If we denote with $-g^2/2$ the corresponding coupling constant, from (\ref{test2}) we have that gauge invariance requires
\be g^2=\hat{e}^2 \label{matching}.\ee
In the following we show that Relation (\ref{matching}) is not
exactly satisfied, but it is broken by very small contributions of
the order of $\epsilon^2$. To prove this statement we observe that
the operator that we are considering has dimension 4 and it may have
both light mode contributions and heavy KK mode contributions
\cite{Randjbar-Daemi:2006gf, Salvio:2006mh}.

We first analyze the light mode contribution $g^2_{lm}$ to $g^2$,
which can be obtained by considering the term
$-e^2\Delta_SA_{\mu}A^{\mu}\sigma^2/2$ in the third line of
(\ref{gaugefixedS}), neglecting the heavy KK modes and integrating
over the extra dimension. Thanks to our exact results of Section
\ref{Bcompletion} we can explicitly compute this contribution and we
find
\be g^2_{lm}=\frac{1}{\sqrt{2\pi}}\frac{e^2\mathcal{M}}{\sqrt{1+\epsilon^2}}.\ee
\begin{figure}[t]
\begin{center}
\begin{picture}(300,140)(0,0)

\Line(-20,180)(0,160)\Text(-20,170)[]{$\omega_1$}
\Line(0,160)(20,180)\Text(20,170)[]{$\omega_1$}
\Line(0,160)(0,125)
\Photon(0,125)(-20,105){2}{5}\Text(-20,115)[]{$V$}
\Photon(0,125)(20,105){2}{5}\Text(20,116)[]{$V$}
\Text(0,100)[]{(a)}

\Line(90,180)(110,160)
\Photon(110,160)(130,180){2}{5}
\Line(110,160)(110,125)
\Photon(110,125)(90,105){2}{5}
\Line(110,125)(130,105)
\Text(110,100)[]{(b)}

\Line(200,160)(220,140)
\Photon(220,140)(200,120){2}{5}
\Line(220,140)(255,140)
\Photon(255,140)(275,120){2}{5}
\Line(255,140)(275,160)
\Text(237,115)[]{(c)}

\Line(-20,70)(0,50)
\Photon(0,50)(20,70){2}{5}
\Photon(0,50)(0,15){2}{5}
\Photon(0,15)(-20,-5){2}{5}
\Line(0,15)(20,-5)
\Text(0,-10)[]{(d)}

\Line(110,50)(130,30)
\Photon(130,30)(110,10){2}{5}
\Photon(130,30)(165,30){2}{5}
\Photon(165,30)(185,10){2}{5}
\Line(165,30)(185,50)
\Text(147,5)[]{(e)}

\end{picture}
\end{center}
\caption{\footnotesize The only types of tree diagrams representing
the heavy mode contribution to the scattering $V,\omega_1
\rightarrow V,\omega_1$. Straight lines are associated to scalar
particles, whereas wavy lines to vector particles. The internal
lines are heavy mode propagators.} \label{diagrams}
\end{figure}
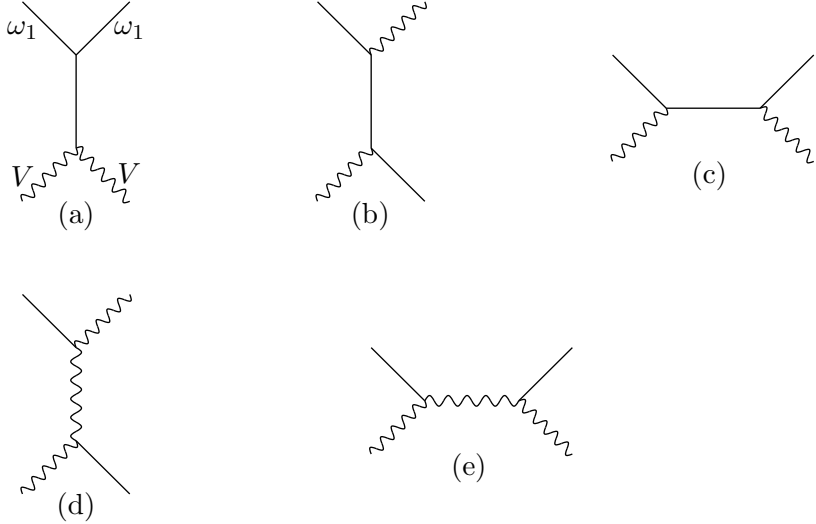

We now pass to the heavy mode contribution to (\ref{4operator}) and
in order to determine it we consider the scattering $V,\omega_1
\rightarrow V,\omega_1$, which represents the scattering between the
Higgs field and the light massive vector. By looking at Action
(\ref{gaugefixedS}) and by using the KK expansions, we find five
types of heavy mode contributions (see Figure \ref{diagrams}).
Diagrams of Type (a) do not contribute as the vertex with two
$\omega_1-$lines and one heavy scalar is proportional to $\int
dy\Delta_S^{-1/2}\chi_{\sigma 1}^2 \chi_{\sigma n}\propto \int dy
\chi_{\sigma 1} \chi_{\sigma n}=0$, where we used $\Delta_S\propto
\chi_{\sigma 1}^2$ and $n=3,5,7...$. Diagrams (b) and (c) are
negligible as they lead to higher dimensional operators in the
effective theory, which involve derivatives. Indeed, one can show
that those diagrams contribute only at the order $\epsilon^4$ to
$g$, if one consistently requires that the momenta of the internal
propagators are much smaller than $\mathcal{M}$. Finally, we
consider Diagrams (d) and (e), which involve a cubic interaction
between two vector bosons and one $\omega_1$. This interaction is
given by the term $-\sqrt2 ve^2\Delta_SA_{\mu}A^{\mu} \sigma$ in the
third line of (\ref{gaugefixedS}). Since this term is proportional
to $v$, Diagrams (d) and (e) do not contribute to $g$ at the order
$1$, but only at the order $\epsilon^2$: we find
\be g_{hm}^2=-\frac{1}{\sqrt{2\pi}}\frac{e^2\mathcal{M}\epsilon^2(1+\epsilon^2)^{-1}}{\sqrt{1+\epsilon^2}+\frac{1}{4}\sqrt{1+\epsilon^2}-\frac{1}{4}},\ee
where $g^2_{hm}$ is the heavy mode contribution to $g^2$. We can now write
\be g^2-\hat{e}^2=g^2_{lm}+g^2_{hm}-\hat{e}^2=-\frac{3}{4}\frac{e^2\mathcal{M}}{\sqrt{2\pi}}\epsilon^2+O\left(\epsilon^4\right).\label{res}\ee
The latter formula shows that the value of $g$ that emerges in the
4D effective theory is not the one required by the gauge invariance.
However, the disagreement is very small because of (\ref{ordereps}).
Indeed, this is a general result: if $\mathcal{O}$ is an observable
quantity in our 4D effective theory and $\mathcal{O}_{GI}$ the
corresponding quantity in a gauge invariant theory, one can
prove\footnote{We checked the validity of Eq. (\ref{effVSgauge}) for
all the interactions by using a method analogous to the one applied
to show (\ref{res}).}
\be \frac{\mathcal{O}}{\mathcal{O}_{GI}}=1+O\left(\epsilon^2\right).\label{effVSgauge}\ee
Eqs. (\ref{res}) and (\ref{effVSgauge}) represent the aforementioned
result: the present model does not admit a 4D effective theory that
can be written as a gauge theory with at most a spontaneous breaking
of the gauge invariance. The reason why this happens is the presence of the weight functions in the
5D lagrangian, which are the only difference with respect to standard KK models. The same property is shared by higher
dimensional gauge theories without fundamental scalars, but with
some weight functions, which diverge when $|y|\rightarrow \infty$
\cite{Shaposhnikov:2001nz}.

As we commented before, the results of this subsection have been derived by neglecting higher order operators in (\ref{test}). In Appendix \ref{HDO} we show
that the higher order operators do not modify these results in the semiclassical approximation.

\subsection{5D versus 4D Higgs mechanism}\label{4Dvs5}

In this subsection we comment about the nature of the Higgs
mechanism in our 5D model. As we pointed out, the main motivation of
the present work is to present a model that relates the SSB of the
gauge symmetry to a chiral asymmetry. However, one can wonder if
the same mechanism may be realized by using a purely 4D language. In
other words, is it possible to reproduce such a result first by
constructing
 the 4D effective theory around an unstable vacuum and then by considering the
SSB in such a theory\footnote{In the following we shall call this approach the {\it 4D effective theory approach to SSB}.}?

The answer to this question is generally negative and, in order to
understand why, we again consider the simple bosonic completion that
we discussed in the present paper. The 4D effective theory approach
to SSB requires to start with the unstable solution $<\phi>=0$ in
the presence of a small\footnote{Here $v$ small means $2v^2\lambda
\ll \mathcal{M}^2$.}  but non vanishing $v$ in the Mexican hat
potential. We observe that this set up can be equivalently achieved
by setting $v=0$ in (\ref{GF}) and (\ref{gaugefixedS}) and
introducing a 5D gauge invariant and tachyonic mass term $+\mu^2
|\phi|^2$ in the lagrangian, where $\mu^2\equiv 2v^2\lambda$.
Therefore, in this case, the light fermion modes are vector-like
because the internal profile $f_0$ of the light vector mode is
constant. Moreover, the 4D effective theory turns out to have an
exact gauge invariance. If one now considers the Higgs mechanism in
such a 4D theory the fermion spectrum will certainly remain chiral
and, by definition, the gauge invariance will be spontaneously - not
explicitly - broken.

This argument clearly shows that the 4D effective theory approach to
SSB does not reproduce the exact values of all the observable
quantities both in the fermionic and in the bosonic sector. However,
we observe that in the bosonic sector this disagreement is very
small because of (\ref{effVSgauge}); therefore it is not very
surprising as we expect the 4D effective theory approach to SSB to
be approximately correct only at energies much smaller than the mass
of the first KK particles\footnote{Indeed, in this limit, such
approach has proved to be correct by considering higher dimensional
scalar theories \cite{Randjbar-Daemi:2006gf}.}. On the other hand,
in the fermionic sector, this disagreement can be important as the
distance $l$ between the left-handed and right-handed branes can
be very large, leading to a non negligible chiral asymmetry.

\section{Non-Abelian Extensions} \label{Extensions}

In this section we comment that a relation between the SSB and the
(low energy) chiral  asymmetry can also be found in the presence of
a non-Abelian gauge group $G$, by generalising in a natural way our
previous analysis. We also discuss the case $G=SU(2)\times U(1)$ and
some difficulties in constructing a realistic model.

\subsection{General non-Abelian gauge groups}

Here we start from a general compact Lie group $G$ with an arbitrary
number of hermitian generators $T^I$, $I=1,...,N$. The gauge field
$A_M$ is a Lie Algebra valued vector field ($A_M=A_M^IT^I$,
$A_M^{\dagger}=A_M$). We introduce a scalar $\phi$ in a non-trivial
representation of $G$ and a scalar potential $V(\phi)$ that triggers
the SSB ($<\phi>\neq 0$). We assume a bosonic action with the same
form as $S_B$ in (\ref{SB}), but with $F_{MN}$ being a non-Abelian
gauge field strength\footnote{Here we assume the vector weight
function $\Delta$ to be universal, namely to be the same function
for all the simple factors in $G$. However, we could also consider
different weight functions for different simple factors.} and $\phi$
in general a multiplet. The covariant derivative of $\phi$ has now
the form
$$D_M\phi=\left(\partial_M+ig_I A_M^I T^I\right)\phi ,$$
where $g_I$ are the gauge constants of $G$ (in general we have more
than one gauge constant). For the sake of simplicity, we take
$<\phi>$ to be constant, which is  a legitimate  set up.

In this more general case, the SSB contribution to the 5D vector boson mass terms  in the 5D lagrangian is
\be-\Delta_S<\phi>^{\dagger}g_IT^Ig_JT^J<\phi>A^I_MA^{JM}\equiv
-\Delta_S T^{IJ}A^I_MA^{JM},\label{general-mass}\ee
where the constant matrix $T^{IJ}$ can be diagonalised
by a constant unitary transformation in the Lie algebra space:
\be \mathcal{A}_M^{\beta}=U_{\beta I}A_M^I, \ee
where $U$ is a unitary matrix. This definition also induces a
redefinition of the generators: $\mathcal{T}^{\beta}=
g_IT^IU^{(-1)}_{I\beta}$, where $U^{(-1)}$ is the inverse of $U$. We
observe that the gauge fields $\mathcal{A}_M^{\beta}$ are not
necessarily real and the generators $\mathcal{T}^{\beta}$ are not
necessarily hermitian.

Also it is possible to add, like in the Abelian case, a gauge fixing
term $\mathcal{L}_{GF}$ in the 5D lagrangian that removes the mixing
between 4D vectors and 4D scalars:
\be
\mathcal{L}_{GF}=-\frac{\Delta}{2}\sum_{\beta}\left|\frac{1}{\Delta}\partial_M\left(\Delta\mathcal{A}^{\beta
M}\right)
-2i\frac{\Delta_S}{\Delta}<\phi>^{\dagger}\left(\mathcal{T}^{\beta}\right)^{\dagger}\Omega\right|^2,\label{non-Ab-GF}\ee
where $\Omega\equiv \phi \,\,- <\phi>$. Gauge fixing
(\ref{non-Ab-GF}) represents the $R_{\xi}$ gauge (for $\xi=1$) in 5D
warped models for general gauge groups and, by generalizing in a
straightforward way Appendix \ref{Rxi}, it is easy to show that it
is indeed a legitimate choice.

With this gauge fixing the EOM of the spin-1 fields are
\be
\frac{1}{\Delta}\partial_M\left(\Delta\partial^M\mathcal{A}^{a\mu}\right)=0,
\quad
\frac{1}{\Delta}\partial_M\left(\Delta\partial^M\mathcal{A}^{\hat{a}\mu}\right)=2
t_{\hat{a}}\frac{\Delta_{S}}{\Delta}\mathcal{A}^{\hat{a}\mu},
\label{EOM-spin-1} \ee
where $\mathcal{A}^{a}_{\mu}$ correspond to the unbroken symmetries ($\mathcal{T}^a<\phi>=0$), $\mathcal{A}^{\hat{a}}_{\mu}$
correspond to the broken symmetries ($\mathcal{T}^{\hat{a}}<\phi>\neq 0$),
\be t_{\hat{a}}\equiv \left|\mathcal{T}^{\hat{a}}<\phi>\right|^2
\label{that}\ee
and in (\ref{EOM-spin-1}) the index $\hat{a}$ is not contracted. It
is now clear that the profiles of the lightest 4D vector fields
coming from $\mathcal{A}^{a}_{\mu}$ are generically constant along
the extra dimension, whereas the profiles of the lightest 4D vector
fields coming from $\mathcal{A}^{\hat{a}}_{\mu}$,
 have generically a non-trivial shape in the broken phase and can be localized on a particular point
of the extra dimensions, say $y=0$, by choosing
suitable weight functions.
Of course, like in the Abelian case, the non-trivial profiles will reduce to constant ones in the unbroken limit $<\phi>\rightarrow 0$.

The 5D fermion field $\psi$ will also belong to a non-trivial
representation of $G$ and, if we assume a fermion action of the
form\footnote{Now we have
$D_M\psi=\left(\partial_M+i\mathcal{A}_M^{\beta}\mathcal{T}^{\beta}_f\right)\psi$,
where $T^{\beta}_f$ are the generators in the fermion
representation.} (\ref{Faction}), we can again localize the lightest
modes of $\psi_L$ and $\psi_R$ on different points of the extra
dimension, say $y=0$ and $y=l$, by choosing a suitable background
domain wall $\varphi$, e.g. Eq. (\ref{phi2}). Analogously to the
Abelian case, the 4D couplings between the light vectors and
fermions can be obtained by integrating over the extra dimensions
the following operators
\be \left[\overline{\psi_L}\gamma^{\mu}\mathcal{A}_{\mu}^a\mathcal{T}_f^a \psi_L+\overline{\psi_R}\gamma^{\mu}\mathcal{A}_{\mu}^a\mathcal{T}_f^a \psi_R+
\overline{\psi_L}\gamma^{\mu}\mathcal{A}_{\mu}^{\hat{a}}\mathcal{T}_f^{\hat{a}} \psi_L+\overline{\psi_R}\gamma^{\mu}\mathcal{A}_{\mu}^{\hat{a}}\mathcal{T}_f^{\hat{a}}
\psi_R\right]_{light},\ee
where the label $''light''$ means that we are selecting only the
light modes in the various KK expansions. Therefore, the fermion
representation of the residual gauge group, with generators
$\mathcal{T}^a$ will certainly be vector-like, whereas the
interactions between fermions and light (but massive) vector bosons
will be chiral. Such chiral asymmetry will reduce to zero in the
unbroken limit for the same reason as it does in the simple Abelian
case.

Here we also observe that we cannot expect the low energy 4D
effective theory to have at most a spontaneous breaking of the gauge
symmetry. This is because the simple 5D warped Higgs model that we
analyzed before (and all the non-Abelian generalizations that reduce
to it via a consistent truncation) represent  explicit
counterexamples.

\subsection{The $SU(2)\times U(1)$ case}

As a particular case, here we consider in more detail the
electroweak case\footnote{The extension to the SM gauge group
$SU(3)\times SU(2) \times U(1)$ is trivial as the $SU(3)$ factor is
not broken.} $G=SU(2)\times U(1)$. Let us denote with $W_M=W_M^I
\frac{\tau^I}{2}$, where $\tau^I$ are the Pauli matrices, and $B_M$
the $SU(2)$ and the $U(1)$ gauge fields respectively. We also choose
the 5D scalar field to transform as $\phi \sim {\bf 2}_{1/2}$ under
$SU(2)\times U(1)$. Therefore, we have
\be D_M\phi =\left(\partial_M +igW_M+i\frac{g'}{2}B_M\right)\phi,
\ee
where $g$ and $g'$ are the gauge constants of $SU(2)$ and $U(1)$
respectively. In this case the bosonic action is
\be S_B=\int d^5
X\left\{-\frac{\Delta}{4}\,\left(W_{MN}W^{MN}+B_{MN}B^{MN}\right)-
\Delta_S \left[\left(D_M\phi\right)^{{\dagger}}D^M\phi +
V(\phi)\right]\right\},\ee
where $W_{MN}$ and $B_{MN}$ are the field strengths of $W_M$ and
$B_M$ respectively and $V(\phi)$ is a scalar potential that triggers
the following VEV
\be <\phi>=\left(\bac 0 \\ v \ea \right), \ee
where $v$ is a real number.

Like in the SM, we now introduce
\bea W_M^{\pm}&=&\frac{1}{\sqrt2}\left(W^1_M\pm i W^2_M\right), \nonumber \\
Z_M&=&\cos\theta \,W_M^3 -\sin\theta \,B_M,\nonumber \\
\gamma_M&=&\sin\theta\, W_M^3+ \cos \theta  \,B_M, \eea
where $\theta$ is defined by $g\sin\theta=g'\cos\theta$. We identify
the lightest modes from $W^{\pm}_{\mu}$, $Z_{\mu}$ and
$\gamma_{\mu}$ with the $W^{\pm}$, $Z$ and photon vector bosons
respectively. The corresponding profiles can be computed through
Eqs. (\ref{EOM-spin-1}) and (\ref{that}); in our case we have
\be t_+=t_-=\frac{1}{4}v^2g^2\,, \qquad t_Z=\frac{1}{4}v^2
\left(g^2+g'^2\right)\,,\label{t-parameters}\ee
where $t_{\pm}$ and $t_Z$ are the parameters defined in (\ref{that})
for $W_{\mu}^{\pm}$ and $Z_{\mu}$ respectively. The relation
$g\sin\theta=g'\cos\theta$ guarantees that the linearized EOM for
$\gamma_{\mu}$ is simply
\be \frac{1}{\Delta}\partial_M\left(\Delta\partial^M
\gamma^{\mu}\right)=0.\ee
Therefore, in the broken phase, we can localize the lightest modes
of $W_{\mu}^{\pm}$ and $Z_{\mu}$ on some point of the extra
dimension, say $y=0$, with a suitable choice of the weight
functions. In the unbroken limit ($v\rightarrow 0 $) these profiles
will go to constants. On the other hand, the lightest mode of
$\gamma_{\mu}$ is generically delocalized and massless both for
$v\neq 0 $ and $v=0$, because the corresponding gauge symmetry is
unbroken.

Now we consider a 5D fermion field\footnote{Here we understand a
flavor index and, therefore, the number of family is generic.}
$\psi$ transforming as $\psi \sim {\bf 2}_{-1/2}$. We can decompose
$\psi$ as
\be \psi=\left(\bac \nu \\ e \ea \right), \ee
and identify the lightest mode of $e$ and $\nu$ with an electron and
a neutrino respectively. Since we have a 5D space-time we
necessarily have both the 4D chiralities: $e=e_L+e_R$ and $\nu=\nu_L
+\nu_R$. If we again localize the left-handed fermions on $y=0$ and
the right-handed ones on $y=l$,  we should be able to reproduce realistic couplings between fermions and $W^{\pm}$: the
interaction between $W^{\pm}$ and the left-handed fermions
%$e_L$ and $\nu_L$
can be achieved by
choosing a suitable value of $g$, whereas the coupling between
$W^{\pm}$
%$e_R$ and $\nu_R$
and the right-handed fermions is suppressed by construction. The
coupling between $\gamma$ and the fermions may also be realistic: we
are free to adjust the coupling between $e$ and $\gamma_{\mu}$ by
properly choosing $g'$ and the interaction between $\nu$ and
$\gamma_{\mu}$ is automatically zero\footnote{This is a consequence
of our choice $\psi\sim {\bf 2}_{-1/2}$, like in the SM.}. The only
interactions between fermions and vector bosons which certainly
cannot be reproduced in this simple framework are those involving
$Z$. Indeed, if we do not modify our set up, the coupling between
$Z$ and the right-handed electron will turn out to be suppressed as
the respective profiles are localized on two different points of the
extra dimension, which we take far away each other. We also observe
that this mismatch is due to the fact that in the SM we have
$g'_{SM}\neq 0$ and so $Z$ interacts non trivially with the
right-handed electron. Therefore, our simple model reproduce the
correct fermion-vector interactions in the limit $g'\rightarrow 0$.

To conclude, the minimal implementation of our mechanism can
reproduce phenomenologically correct interactions between the
fermions and $\{W^{\pm},\,\gamma\}$, but is not general enough to be
realistic. This is true not only because the correct interactions of
$Z$ cannot be reproduced, but also because it seems difficult, at
least in this simple set up, to achieve the correct fermion and
vector boson masses. Indeed, we could think to introduce some Yukawa
interactions in the 5D lagrangian in order to obtain a realistic
fermion mass spectrum, but then we would probably introduce a
non-universality in the gauge interactions of different families.
Moreover, the spectrum of the vector bosons is also problematic
because the value of the $t$-parameters appearing in
(\ref{t-parameters}) are not necessarily proportional to the masses
of the vector bosons (like in the SM). Indeed, these masses must be
computed by solving the Eqs (\ref{EOM-spin-1}) and therefore will
depend in general on the shape of the wave functions.

However, we consider the present discussion interesting as we found
a relation between the chiral asymmetry and the Higgs mechanism in a
model that resembles the electroweak theory in the low energy limit.
We plan to study more general and possibly realistic implementation
of our mechanism in a future phenomenological extension of the
present theoretical work.

\section{Conclusions and Outlook}\label{conclusions}

In this paper we have proposed a mechanism that relates the chiral
asymmetry to the 5D Higgs mechanism, which generates masses for the
low energy degrees of freedom. To illustrate the basic idea we have
analyzed a 5D fermion and, as a bosonic completion, we have
considered a simple 5D {\it warped} Higgs model. Such mechanism
exploits the fact that the 4D gauge field profile along the extra dimension is generally constant if the gauge symmetry is unbroken, but
it can be peaked around some point, for example $y=0$, in the broken case.
This point also represents the 4D world where fermions with a given
chirality live, whereas the other chirality can be localized on
another point $y=l$, by means of a domain wall configuration.

A specific feature of our mechanism is that, in the case $v\neq 0$,
the bosonic action of the 4D effective theory generically cannot be
written as a gauge invariant action with at most a spontaneous
breaking of the gauge symmetry. This statement has been proved by
choosing some specific values of the weight functions, which may
have their origin in warped compactification of theories of gravity
or supergravity. Indeed, for these specific values, all the bosonic
profiles along the extra dimension are simply 1D harmonic oscillator
wave functions and, therefore, many observable quantities can be
explicitly computed. Then, we have considered two possible physical
definitions of the 4D gauge constant, which must coincide in a gauge
invariant theory. The first definition can be obtained from the
decay amplitude of the physical Higgs field $\omega_1$ into two
light vector bosons $V$, whereas the second one comes from the
scattering $\omega_1,V\rightarrow \omega_1,V$. These coupling
constants turned out to be different, even if their relative
difference is a very small quantity, of the order of the square of
the ratio between the {\it electroweak} scale and the lightest heavy
mode mass. Remarkably, the 4D effective theory becomes gauge
invariant, at least at the semiclassical level, as $v\rightarrow 0$
that is exactly the limit in which the chiral asymmetry disappears.

As a consequence of the aforementioned results, the Higgs mechanism,
which we triggered in order to achieve our purpose, cannot be
described by a purely 4D method, that we called 4D effective theory
approach to SSB and defined in Subsection \ref{4Dvs5}. Indeed, the
latter method is expected to be correct at the leading order in
$\epsilon$ in standard KK theories, but in our model breaks down
because of the presence of an additional parameter, $l$. By taking
into account $\epsilon^2$ corrections, such a method turned out to
be incorrect even without fermions. This is not very surprising,
because the 4D effective theory is expected to be correct only at
energies much smaller than the heavy KK mode mass, but is still
interesting because in our model the $\epsilon^2$ corrections are
the first non-trivial corrections.

Moreover, we  commented that a relation between the chiral asymmetry
and the Higgs mechanism can also emerge in non-Abelian
generalizations of the warped Higgs model with fermions, which we
have previously analyzed. We showed that the unbroken directions in
the Lie algebra space correspond to vector-like interactions in the
low energy theory, whereas the remaining directions present a chiral
asymmetry. As an explicit non-Abelian example we considered the
electroweak group $SU(2) \times U(1)$ and we obtained, by using a
simple set up, a low energy limit that resembles - but does not
entirely reproduce - the standard electroweak theory. Some of the
reasons why the simple implementation that we considered is not
realistic are the fact that the correct interactions of Z are not
reproduced and it seems difficult to obtain realistic fermion and
vector boson masses.

An interesting development of the present work can be the extension
of the simple implementation presented in this paper to a theory
which exactly reduces to the SM at low energies. For example,
extensions of this type might be the introduction of more than two
branes, general Yukawa couplings and general weight functions in the
higher dimensional model. Moreover, it would be nice to have an
embedding of this type of models in a more fundamental theory which
includes gravity. Indeed, this may lead to a dynamical origin of the
domain wall configuration $\varphi$ and the weight functions.
Finally, another interesting direction is the complete calculation
of the heavy mode contribution to the 4D effective theory, including
the effect of the heavy fermion fields: an exact cancellation of the
chiral anomaly may happen as in previous works
\cite{Callan:1984sa,Kaplan:1995pe} and some observational effects
could emerge \cite{Khlebnikov:1987zg}.

\newpage%vspace{0.2cm}

{\bf Acknowledgments.}   We would especially like to thank A. Boyarsky, Y. Burnier, R. Rattazzi, O. Ruchayskiy, A. Wulzer and K. Zuleta
for interesting and useful communications.
This work was supported by the Tomalla Foundation.
\vspace{1.5cm}

%\newpage
\appendix
{\Large \bf Appendix}
%\phantom{}\vskip 1cm  {\Large \bf $R_{\xi}$ Gauges in 5D Warped Models}\vskip 1cm
\section{$R_{\xi}$ Gauges in 5D Warped Models}\label{Rxi}
%\hspace{-0.6cm}
In this appendix we show that the following class\footnote{Since we
can do so without much expense, here we keep a generic value of the constant $\xi$.}  of gauge fixing terms
\be \mathcal{L}_{GF}=-\frac{\Delta}{2\xi}\left[\frac{1}{\Delta}\partial_M\left(\Delta
A^M\right)+\sqrt{2}ev\xi\frac{\Delta_S}{\Delta}\eta\right]^2, \ee
represents possible gauge fixings. In analogy with the 4D case, we
refer to them as  $R_{\xi}$ gauges. More precisely, here we prove
that, for a given initial configuration $\mathcal{G}_i$ of the gauge
function $\mathcal{G}\equiv (1/\Delta)\partial_M(\Delta
A^M)+\sqrt{2}ev\xi(\Delta_S/\Delta)\eta$ and for a given arbitrary
space time function $\varepsilon(X)$, we can always find a gauge
transformation of the form (\ref{gaugetransf}) such that
$\mathcal{G}_i\rightarrow \mathcal{G}_i+\varepsilon$. In the rest we
assume $\varepsilon$ to be infinitesimal as the general case can be
addressed by considering an infinite number of infinitesimal
transformations.

Therefore, we have to explicitly find a solution to the following equation
\be -\frac{1}{\Delta}\partial_M\left(\Delta\partial^M\alpha \right)+ e^2K\alpha=\varepsilon,\label{GFeq}\ee
where $K\equiv 2 v\xi
\frac{\Delta_S}{\Delta}\left(v+\sigma/\sqrt2\right)$. In
(\ref{GFeq}) we used $\mbox{Im}\left(e^{ie\alpha}\phi\right)=
e\alpha\left(v+\sigma/\sqrt2 \right)+\eta/\sqrt2$, which is valid
for an infinitesimal value of $\alpha$. We can find a solution to Eq. (\ref{GFeq}) by using the perturbation theory with respect to $e^2$;
a  non-perturbative treatment of Eq. (\ref{GFeq}) is beyond our
purposes since we always used a perturbative approach in this
paper\footnote{We thank Riccardo Rattazzi for a discussion on this
issue.}.

We start by considering the unperturbed ($e=0$) equation:
\be -\frac{1}{\Delta}\partial_M\left(\Delta\partial^M\alpha_0 \right)=\varepsilon. \label{GFeq0}\ee
A solution $\alpha_0$ to this equation can be written in terms of the Green's function $G_0(X,X')$ for the operator
$-\frac{1}{\Delta}\partial_M\left(\Delta\partial^M\right.$:
\be
G_0(X,X')=\sum_n\int\frac{d^4q}{(2\pi)^4}\frac{1}{q^2+\lambda_n}e^{iq(x-x')}\psi_n(y)\psi_n^*(y'),\ee
where $\{\psi_n\}$ is a complete set of eigenfunctions of the operator $-\Delta^{-1}\partial_y(\Delta\partial_y$ and $\lambda_n$ the corresponding
eigenvalues. We have already studied this basis in Subsection \ref{spin-1}:
$$\psi_n=\lim_{v \rightarrow 0} f_n \qquad \mbox{and} \qquad \lambda_n=\lim_{v \rightarrow 0}M^2_n.$$
A solution to Eq. (\ref{GFeq0}) can be now written as follows
\be \alpha_0=\int d^5X_1 G_0(X,X_1)\varepsilon(X_1), \label{alpha0}\ee
whereas a complete solution to Eq. (\ref{GFeq}) can be expressed in
terms of a Taylor series with respect to $e^2$, that is
$\alpha=\sum_{j=0}^{+\infty}\alpha_j$, where $\alpha_0$ is given in (\ref{alpha0}) and
\bea  \alpha_1(X)&=&-e^2\int d^5X_1 G_0(X,X_1)K(X_1)\int d^5X_2 G_0(X_1,X_2) \varepsilon (X_2), \nonumber \\
  \dots, \alpha_j(X)&=&(-1)^je^{2j}\int d^5X_1 G_0(X,X_1)K(X_1)\int d^5 X_2G_0(X_1,X_2)K(X_2) \times  ...\times  \nonumber \\
&&\int d^5X_j G_0(X_{j-1},X_j)K(X_j)\int d^5X_{j+1} G_0(X_j,X_{j+1})\varepsilon(X_{j+1}). \nonumber \eea

\section{Higher Dimensional Operators}\label{HDO}

Here we analyze the higher dimensional gauge invariant operators in (\ref{test}) that could modify (after SSB) the operators $V_{\mu}V^{\mu}$,
$V_{\mu}V^{\mu}\omega_1$ and $V_{\mu}V^{\mu}\omega_1^2$. Indeed, other operators cannot ruin the argument given in Section \ref{4Deff}
that leads to a small gauge symmetry breaking in the effective theory.

Such higher dimensional operators should contain a term with no
derivatives and where $V_{\mu}$ appears in the form
$V_{\mu}V^{\mu}$. We also observe that the only way to construct
gauge invariant operators which contain $V_{\mu}$ is through
$D_{\mu}\omega$ and $V_{\mu \nu}$, but it is impossible to construct
an operator without derivatives if we use $V_{\mu \nu}$. Therefore,
we should start from the following type of gauge invariant
operators:
\be \left(\omega^{\dagger}
\omega\right)^p\left(D_{\mu}\omega\right)^{\dagger}D^{\mu}\omega,
\label{H-operators}\ee
where $p=1,2,3,...$. Indeed, if we started from an operator which contains more than two covariant derivatives $D_{\mu}\omega$
and then extracted a term where $V_{\mu}$ appears in the form
$V_{\mu}V^{\mu}$, we would necessarily have some derivatives as well.

Let us show that (\ref{H-operators}) cannot be derived from our 5D
theory, whose action is given in (\ref{gaugefixedS}). To construct
(\ref{H-operators}), among other things, we need the following
operator:
\be \omega_1^{2p}\partial_{\mu}\omega_1\partial^{\mu}\omega_1,\label{H2-operators}\ee
which follows from (\ref{H-operators}) when we select $\omega_1$ in the expansion $\omega= \hat{v}+(\omega_1+i\omega_2)/\sqrt2$, and $\partial_{\mu}\omega_1$ in the expansion
$D_{\mu}\omega=\left(\partial_{\mu}+i\hat{e}V_{\mu}\right)\left[\hat{v}+(\omega_1+i\omega_2)/\sqrt2\right]$. Remember also $\omega_1\equiv \sigma_1$.

We prove now that operators of the form (\ref{H2-operators}) can
appear in our model only through loop corrections, which we do not
include in the present analysis. First notice that, in order to
construct a tree-level contribution to (\ref{H2-operators}), we need
at least one vertex containing only one heavy line (this is because
in tree-diagrams we must have some vertices where an internal line
ends). Vertices with only one heavy line and light lines of the
$\sigma_1$-type only come from the following
interactions\footnote{Though (\ref{suit.inter}) does not contain
derivatives as in (\ref{H2-operators}), they arise after expanding
over the momentum.} in (\ref{gaugefixedS}):
\be \int d^5X \left( -\sqrt2 v \lambda \Delta_S \sigma^3 -\frac{\lambda}{4}\Delta_S\sigma^4\right). \label{suit.inter}\ee
 The first and second terms in (\ref{suit.inter}) lead to two classes of 4D interactions of the required type,
 whose coupling constants are respectively proportional to
\be \int dy \Delta_S^{-1/2}\chi_{\sigma 1}^2 \chi_{\sigma n},\quad \mbox{and }\quad \int dy \Delta_S^{-1}\chi_{\sigma 1}^3 \chi_{\sigma n},\ee
where $n>1$. Now, by using $\chi_{\sigma 1}\propto \Delta_S^{1/2}$
and the orthogonality between $\chi_{\sigma 1 }$ and $\chi_{\sigma
n}$, with $n>1$, we have
\be \int dy \Delta_S^{-1/2}\chi_{\sigma 1}^2 \chi_{\sigma n}\propto \int dy \chi_{\sigma 1}\chi_{\sigma n}=0\ee
and
\be \int dy \Delta_S^{-1}\chi_{\sigma 1}^3 \chi_{\sigma n}\propto \int dy \chi_{\sigma 1}\chi_{\sigma n}=0. \ee
Therefore, we cannot construct (at the semiclassical level) a gauge
invariant operator of the form (\ref{H-operators}) in our model. We
conclude that the argument provided in Section \ref{4Deff} is valid
even if we take into account higher order operators in (\ref{test}).

\end{document}